\documentclass[aps,twocolumn,pra]{revtex4-1}%
\usepackage{amsfonts}
\usepackage{amsmath}
\usepackage{amssymb}
\usepackage{graphicx}
\usepackage{times}
\usepackage{dsfont}
\usepackage{bm}
\usepackage[colorlinks=true,citecolor=blue,urlcolor=blue]{hyperref}%
\providecommand{\U}[1]{\protect\rule{.1in}{.1in}}
\begin{document}
\title{Atomic Interferometry with Spin-Orbit-Coupled Spin-1 Condensates}
\author{Renfei Zheng$^{1,4}$}
\thanks{These authors contributed equally.}
\author{Junying Wu$^{1}$}
\thanks{These authors contributed equally.}
\author{Josep Cabedo$^{2}$}
\author{Alessio Celi$^{2}$}
\author{Zhihao Lan$^{3}$}
\author{Weiping Zhang$^{5,6,7,8,9}$}
\author{Lu Zhou$^{1}$}
\email{lzhou@phy.ecnu.edu.cn}
\affiliation{$^{1}${State Key Laboratory of Precision Spectroscopy,
School of Physics, 
East China Normal University,
Shanghai 200241, China}}
\affiliation{$^{2}${Departament de Física, Universitat Aut\`onoma de Barcelona,
E-08193 Bellaterra, Spain}}
\affiliation{$^{3}${College of Physical Sciences and Engineering, Mohammed VI Polytechnic University, Ben Guerir, 43150, Morocco}}
\affiliation{$^{4}${School of Physics, Hefei University of Technology, Hefei 230009, China}}
\affiliation{$^{5}${School of Physics and Astronomy, Shanghai Jiao Tong University,
Shanghai 200240, China}}
\affiliation{$^{6}${Tsung-Dao Lee institute, Shanghai Jiao Tong University, Shanghai
200240, China}}
\affiliation{$^{7}${Shanghai Branch, Hefei National Laboratory, Shanghai 201315, China}}
\affiliation{$^{8}${Shanghai Research Center for Quantum Sciences, Shanghai 201315, China} }
\affiliation{$^{9}${Collaborative Innovation Center of Extreme Optics, Shanxi University,
Taiyuan, Shanxi 030006, China} }

\begin{abstract}
We propose and analyze a quantum interferometry scheme based on a Raman-dressed Bose gas with spin-orbit coupling. In this system, the atom-light coupling mixes spin and momentum degrees of freedom, giving rise, in the low-energy regime, to an effective spinor condensate whose spin-mixing interaction can be tuned independently of the atomic density. This controllability enables a separation between state preparation and phase imprinting, and provides a natural route to echo-type protocols based on effective time reversal. Within this framework, critical regimes of the effective spinor Hamiltonian can be used to generate entanglement and enhance interferometric sensitivity beyond the standard quantum limit. In addition, the spin-momentum locking of the dressed modes gives access to spatial density modulations that provide an alternative readout of the interferometric phase. In particular, phase information can be extracted from the displacement of spin-orbit-induced density stripes even when conventional spin observables are insensitive within the effective spinor description. Our results identify Raman-dressed spinor gases as a flexible platform for nonlinear atomic interferometry, combining controllable spin-mixing dynamics with spatially resolved phase readout.
\end{abstract}
\maketitle

\section{introduction} % (fold)
\label{sec:introduction}

Spinor Bose-Einstein condensates (BECs) constitute a natural platform for atom interferometry, as their intrinsic interactions enable the generation of squeezing and entanglement through spin-exchange collisions \cite{Stamper-Kurn2013,Pu2000,Duan2000,Hamley2012,Luo2017}. These interactions give rise to nonlinear spin-mixing dynamics that can be exploited for interferometric protocols beyond the standard quantum limit, including implementations of SU(1,1) interferometry and related nonlinear schemes \cite{Hudelist2014,Gabbrielli2015,Linnemann2016,Wrubel2018,Qu2020,Law1998,Liu2022}. More recently, critical behavior in such systems has been identified as a resource for quantum metrology, as the enhanced fluctuations near quantum phase transitions can be harnessed to improve parameter estimation \cite{zanardi2008,macieszczak2016,garbe2020,chu2021,Ding2022,ilias2022,Garbe_2022,Gietka2022understanding,Aybar2022criticalquantum,guan2021,wang2024,xue2026,zhouPRR}. In addition, recent work has shown that information about the system’s phase can be extracted from spin-mixing dynamics, providing experimental access to dynamical quantum phase transitions \cite{Meyer-Hoppe2023,Austin-Harris2025}. Despite these promising features, implementing interferometric protocols in spinor condensates typically requires a careful balance between interaction dynamics and measurement strategies, and can be constrained by the available control over the spin-mixing interactions and by the set of accessible observables. 
% \textcolor{red}{
    For instance to make independent phase imprinting, 
    people in \cite{Liu2022} have to quench the quadratic Zeeman energy to a sufficiently large value to suppress the spin mixing dynamics.    
% }
In this work, we propose a scheme that addresses these aspects by exploiting 
% a form of synthetic spin-orbit-coupling (SOC) in Raman-dressed quantum gases, where both 
Raman light to control both
the effective interactions and the readout 
% can be engineered 
in a flexible way.

% On the other hand, 
Raman light induces a synthetic
spin-orbit coupling (SOC) in ultracold atoms \cite{Lin2009,Lin2011}
% has emerged as 
and is
a versatile tool to engineer synthetic gauge fields 
% using atom–light interactions
and many-body phases
 \cite{Dalibard2011,Goldman_2014,Zhai_2015,Zhang2016}. 
 % In particular, Raman-dressed spinor gases  exhibit a rich phenomenology, including modified interaction properties \cite{Williams2012}, stripe phases with supersolid-like character \cite{Li2017,Putra2020,Chisholm2026}, and the realization of more exotic effective models \cite{Frolian2022}. A key feature of SOC systems is the coupling between the internal spin and the center-of-mass motion, which links spatial dynamics to spin degrees of freedom. 
 % This coupling 
The coupling of internal and external degrees of freedom
 has been also exploited in interferometric settings, where spatial phase accumulation can be mapped onto spin observables, enabling applications such as force sensing and the simulation of gauge-field effects \cite{Anderson2011,Jacob2007,Osterloh2005}. Interferometric protocols based on SOC condensates have also been explored, including Stückelberg interferometry \cite{Olson2017,Liang2020}, and proposals for enhanced precision measurements \cite{Gietka2023}.

In this work, we propose interferometric protocols based on a spin-1 Bose–Einstein condensate with 
% spin–orbit coupling induced by Raman dressing
SOC. In the low-energy regime, the system can be described by an effective spinor model whose spin-mixing interaction can be tuned independently of the atomic density, providing direct control over the nonlinear dynamics \cite{CabedoPRA2021,Cabedo2021}. 
% An key feature of this platform is the coupling between internal (spin) and external (momentum) degrees of freedom, which 
The coupling between spin and momentum
enables the interferometric phase to be encoded not only in spin observables but also in spatial density modulations. Building on these properties, we analyze interferometric schemes based on spin-mixing dynamics and show how the tunability of the interaction can be used to control the different stages of the protocol. In particular, we consider a sequential protocol in which the interaction can be effectively suppressed during phase imprinting, and an echo-based protocol enabled by interaction reversal. We further discuss how critical regimes of the effective Hamiltonian can be exploited to generate entanglement and enhance the sensitivity of the scheme.

The paper is organized as follows. In Sec.~\ref{sec:brief_summary}, we review the effective spinor description of the Raman-dressed system and highlight the tunability of the spin-mixing interaction, which plays a central role in our proposal. In Sec.~\ref{sec:interfere_scheme}, we consider interferometric protocols based on spin-mixing dynamics and discuss how the tunability of the interactions in our system can be used to control the different stages of the interferometric sequence. In particular, we analyze both a sequential protocol (Sec.~\ref{sec:interfere_scheme}A) and an echo-based protocol enabled by interaction reversal (Sec.~\ref{sec:interfere_scheme}B), and discuss how critical regimes of the effective Hamiltonian can be incorporated to enhance the sensitivity of the scheme. In Sec.~\ref{sec:stripe_phase}, we explore how SOC-induced density modulations can be exploited to provide an alternative readout of the interferometric phase, particularly in regimes where conventional spin-based observables become ineffective. Finally, we summarize our findings and discuss future directions in Sec.~\ref{sec:conclusion}.
% =============================

% section review    ===========
\section{
Effective spinor description of a spin-orbit-coupled BEC}
\label{sec:brief_summary}

\begin{figure}
[htb]\centering
\includegraphics[width=3 in]{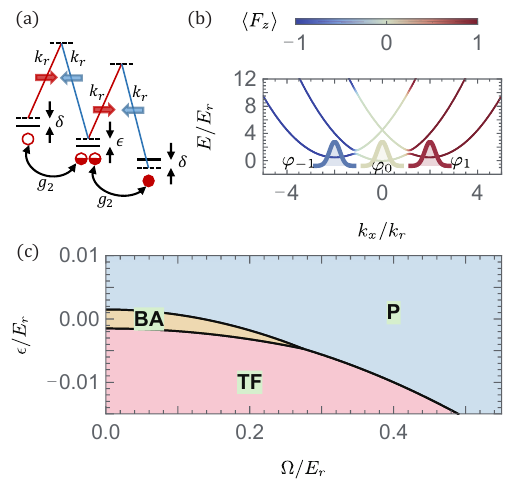} 
\caption{
Effective spinor description of a Raman-dressed spin-1 BEC.
(a) Raman coupling scheme within the $F=1$ manifold. The Raman lasers couple adjacent Zeeman states while transferring momentum $2\hbar k_r$. The diagram also indicates the spin-changing collision process, proportional to $g_2$, in which two atoms in the $m=0$ state are converted into a correlated pair in the $m=\pm1$ states, and vice versa. The linear and quadratic detunings from Raman resonance are denoted by $\delta$ and $\epsilon$, respectively.
(b) Dispersion bands of Hamiltonian~(\ref{eq:H_SOC}) for $\Omega=0.5E_r$, $\delta=0$, and $\epsilon=0.5E_r$. The color scale indicates the expectation value of $F_z$ for the corresponding band states. In the weak Raman coupling regime, the lowest band exhibits a triple-well structure. The low-energy dynamics can then be described in terms of three dressed modes,
$\bm{\varphi}=(\varphi_1,\varphi_0,\varphi_{-1})^T$,
localized around the three minima.
(c) Ground-state phase diagram of the effective spinor Hamiltonian~(\ref{eq:H_eff}) for $g_2 < 0$, shown as a function of the Raman coupling $\Omega$ and the quadratic Zeeman detuning $\epsilon$. The polar (P), twin-Fock (TF), and broken-axisymmetry (BA) phases meet at a tricritical point located at $\Omega_c = 4 E_r\sqrt{|g_2|/g_0}$, which separates the ferromagnetic ($\lambda < 0$, for $\Omega < \Omega_c$) and antiferromagnetic ($\lambda > 0$, for $\Omega > \Omega_c$) regimes of the effective Hamiltonian.}
\label{fig:phase_diagram}
\end{figure}

In this section, we briefly review how a Raman-dressed spin-1 BEC can be mapped onto an effective spinor system in the low-energy regime, following Ref.~\cite{CabedoPRA2021}. As illustrated in Fig.~\ref{fig:phase_diagram}(a), we consider a spin-1 BEC of $N$ atoms in the $\vert F = 1, m = 0, \pm 1 \rangle$ hyperfine manifold, subject to Raman coupling generated by two counter-propagating laser beams, which impart a momentum transfer and induce spin-orbit coupling in the system. Spin-orbit-coupled spin-1 condensates were first theoretically explored in \cite{lan2014} and later realized experimentally in \cite{Campbell2016}.

The corresponding single-particle Hamiltonian can be written as
\begin{equation}
    \mathcal{H}_{\text{SOC}} = \frac{\hbar^2}{2m}
    (\bm{k}-2 k_r F_z \bm{e}_x)^2
    +
    \frac{\Omega}{\sqrt{2}} F_x
    +
    \delta F_z
    +
    \epsilon F_z^2,
\label{eq:H_SOC}
\end{equation}
where $F_{x,y,z}$ are the spin-1 matrices. Here, $k_r$ denotes the recoil momentum associated with the Raman process, $\Omega$ is the Raman coupling strength, and $\delta$ and $\epsilon$ are the linear and quadratic detunings from Raman resonance, which can be controlled experimentally, for instance via microwave dressing \cite{Gerbier2006,Leslie2009}.

% figure phase ffffffffffffffff
% fffffffffffffffffffffffffffff

% \textcolor{blue}{say dispersion relation}

The Hamiltonian in Eq.~(\ref{eq:H_SOC}) can be diagonalized in the quasimomentum $k_x$ basis, yielding three dispersion branches, as shown in Fig.~\ref{fig:phase_diagram}(b), where energies are expressed in units of the recoil energy $E_r = \hbar^2 k_r^2 / 2m$. 
In the weak Raman coupling regime, the lowest dispersion branch exhibits three well-separated minima. For sufficiently small condensates and strong confinement, the low-energy dynamics remains predominantly localized tightly around these minima, allowing for an effective description in terms of three dressed modes,
$\bm{\varphi} = (\varphi_1, \varphi_0, \varphi_{-1})^T$
% \textcolor{red}{
    composed of products of bare spin states and corresponding spatial wavefunctions \cite{CabedoPRA2021}.
    The validity of three-mode truncation has been verified via comparing the Gross--Pitaevskii equation (GPE) propagation results with that obtained from the three mode Hamiltonian, good agreements are found in the parameter regime considered throughout the present work.
% }

We denote by $b_{j=0,\pm 1}^\dagger$ the creation operator associated with the dressed mode $\varphi_j$, and by $N_j = b_j^\dagger b_j$ the corresponding occupation number. In this basis, we introduce the collective spin operators
$L_i = b_\alpha^\dagger F_i^{\alpha\beta} b_\beta$ , $(i = x,y,z),
$ and $
L_{zz} = b_\alpha^\dagger (F_z^2)^{\alpha\beta} b_\beta
$, where repeated indices $\alpha,\beta$ are summed over the spin components. Henceforth, we focus on the subspace of zero magnetization, which is the relevant regime for the present work. Within the three-mode truncation, the many-body dynamics of the dressed gas can then be effectively described by the effective Hamiltonian
\begin{equation}
    H = \lambda \frac{L^2}{2 N}
    +
    \tilde{\epsilon} L_{zz}.
\label{eq:H_eff}
\end{equation}
Here, $\lambda$ and $\tilde{\epsilon}$ denote the effective spin-mixing interaction strength and quadratic Zeeman energy, respectively, given by
\begin{equation}
    \lambda = \left(g_2 + g_0 \frac{\Omega^2}{16 E_r^2}\right) \bar{n},
    \qquad
    \tilde{\epsilon} = \epsilon + \frac{\Omega^2}{16 E_r^2},
\label{eq:spinor_para}
\end{equation}
where $g_0$ and $g_2$ characterize the spin-symmetric and spin-dependent interactions, and $\bar{n}$ is the mean density of the gas.

The effective Hamiltonian in Eq.~(\ref{eq:H_eff}) has the same structure as that of a spinor BEC subject to a quadratic Zeeman shift $\tilde{\epsilon}$ \cite{Law1998}. A key feature of the present platform, however, is that the spin-mixing interaction $\lambda$ can be tuned independently via the Raman coupling, as shown in Eqs.~(\ref{eq:spinor_para}). 
% In the following, we focus on the case of 
For a ferromagnetic condensate with bare spin-dependent interaction $g_2 < 0$, as realized for instance in $^{87}$Rb, $\lambda$ can be tuned across zero and even change sign by adjusting $\Omega$. The resulting ground-state phase diagram, shown in Fig.~\ref{fig:phase_diagram}(c) as a function of $\Omega$ and $\epsilon$, reflects this tunability. In the thermodynamic limit, the system exhibits the familiar polar (P), twin-Fock (TF), and broken-axisymmetry (BA) phases of the spin-1 spinor gas. In the P and TF phases, one has $\rho_0 = \langle b_0^\dagger b_0 \rangle / N = 1$ and $0$, respectively, while the BA phase is characterized by a transverse magnetization that breaks the axial SO(2) symmetry \cite{Sadler2006,Murata2007}.

While the phase diagram itself closely follows that of conventional spinor condensates, the independent control of $\lambda$ enables a wider range of dynamical protocols, which will be central to the interferometric schemes discussed below. Importantly, the dressed modes inherit a nontrivial quasimomentum structure from SOC, which translates into spatial density modulations in the condensate. As a result, the different phases of the effective model acquire distinct spatial structures. For instance, the broken-axisymmetry phase corresponds to a stripe state, while modulated patterns can persist more generally across the low-energy spectrum \cite{CabedoPRA2021, Cabedo2021}.

% section interfere ===========
\section{Interferometric protocols with tunable spin-mixing interactions} % (fold)
\label{sec:interfere_scheme}

Building on the tunability of the spin-mixing interaction discussed above, we now consider interferometric protocols based on spin-mixing dynamics in this platform. While the underlying schemes are closely related to those developed for conventional spinor BECs, the independent control of the interaction strength provides additional flexibility in shaping the different stages of the interferometric sequence. In particular, it allows one to manipulate the interaction during phase imprinting or even reverse its sign, enabling distinct strategies for interferometric operation. In the following, we analyze two representative cases: a sequential protocol, in which the interaction can be effectively suppressed during the phase-imprinting stage, and an echo-based protocol, in which the interaction is reversed to implement an effective time-reversal dynamics.

\subsection{Sequential interferometric protocol}

We begin by considering a sequential interferometric protocol based on the effective spinor Hamiltonian in Eq.~(\ref{eq:H_eff}).
% , following schemes previously developed for spinor BECs \cite{Niezgoda_2019,Kajtoch2018} and illustrated schematically in Fig.~\ref{fig:scheme}. 
The protocol consists of three main stages:

(i) \emph{State preparation.} An initial state $\vert \psi_0 \rangle$ evolves under the Hamiltonian $H$ for a time $t$, generating an entangled state $\vert \psi(t) \rangle = e^{-i H t} \vert \psi_0 \rangle $, which serves as the input state for the interferometer.

(ii) \emph{Phase imprinting.} The system undergoes a unitary evolution $e^{-i \theta \Lambda}$, during which the parameter $\theta$ to be estimated is imprinted. The generator $\Lambda$ depends on the physical quantity being measured; for instance, for magnetic-field sensing via the linear Zeeman effect, one has $\Lambda = L_z$.

(iii) \emph{Measurement.} Finally, a measurement is performed on the evolved state, optionally preceded by a unitary operation $U$. In the present discussion, we take $U = \mathds{1}$.

% figure scheme fffffffffffffff
% \begin{figure}
% [htb]\centering
% \includegraphics[width=3 in]{fig/fig_scheme.pdf} 
% \caption{
% Schematic illustration of the sequential and echo interferometric protocol.
% }
% \label{fig:scheme}
% \end{figure}
% fffffffffffffffffffffffffffff

The performance of the interferometric protocol can be characterized by the quantum Fisher information (QFI), which sets the ultimate precision bound through the quantum Cramér–Rao inequality $\Delta \theta \ge 1/\sqrt{\mathcal{I}_\theta}$. After the state-preparation stage, the system is in the state $\vert \psi(t) \rangle = e^{-iHt}\vert \psi_0 \rangle$. The phase $\theta$ is then imprinted through the unitary transformation $e^{-i\theta\Lambda}$, leading to the state $\vert \psi_\theta \rangle = e^{-i\theta\Lambda}\vert \psi(t) \rangle$. For pure states and unitary encoding, the QFI takes the form $\mathcal{I}_\theta = 4\,(\Delta \Lambda)^2_{\vert \psi(t) \rangle}$, where $(\Delta \Lambda)^2_{\vert \psi(t) \rangle}$ denotes the variance of the generator $\Lambda$ evaluated on the state prior to phase imprinting.

In the present platform, the tunability of the spin-mixing interaction provides additional flexibility in implementing this protocol. In particular, the interaction strength $\lambda$ can be adjusted independently via the Raman coupling, allowing one to effectively suppress the nonlinear dynamics during the phase-imprinting stage. That is, the system can be evolved under the interacting Hamiltonian during state preparation, while the interaction can be tuned close to zero when the phase $\theta$ is imprinted. This separation of roles can be especially advantageous for the detection of weak signals, where unwanted nonlinear evolution during phase accumulation can degrade sensitivity.

In the following, we consider alternative interferometric strategies that further exploit the tunability of the interaction in this platform.

% criticality   ===============
\subsection{Echo interferometric protocol}
\label{sec:critical}

% In this section, we consider
The sequential protocol can be extended to
an echo-type interferometric protocol, 
% as shown in Fig.~\ref{fig:scheme},
in which the phase imprinting is embedded between a forward and a backward evolution under the interacting Hamiltonian. The resulting state takes the form
$\vert \psi(\theta) \rangle = e^{iHt} e^{-i \theta \Lambda} e^{-iHt} \vert \psi_0 \rangle$, which resembles a time-reversal (echo) sequence \cite{Davis2016,Florian2016,Tommaso2016}. 
By measuring the fidelity
$\vert \langle \psi(\theta) \vert \psi_0 \rangle \vert^2$,
the QFI $\mathcal{I}_\theta$ can be extracted.
Implementing such a protocol requires reversing the sign of the Hamiltonian, and in particular of the spin-mixing interaction parameter $\lambda$.

This requirement highlights a key advantage of the 
% present platform. 
Raman-dressed spinor condensate.
In a conventional spinor condensate, the sign of the spin-dependent interaction is fixed by the atomic species and cannot be easily reversed. As a result, implementing an exact time-reversal of the dynamics is experimentally challenging. To the best of our knowledge, effective sign reversal of the interaction has only been demonstrated under restrictive conditions such as the non-depletion regime \cite{Linnemann2016}. In contrast, in the SOC-based realization of the effective Hamiltonian (\ref{eq:H_eff}), the interaction parameter $\lambda$ can be tuned in magnitude and, for systems with a ferromagnetic bare spin-dependent interaction ($g_2 < 0$), also in sign by adjusting the Raman coupling $\Omega$ (see Fig.~\ref{fig:phase_diagram}(c)). This enables a practical implementation of echo-type protocols based on effective time reversal.

To exploit this protocol for enhanced metrology, one needs to prepare input states with large quantum fluctuations in the generator $\Lambda$. 
% \textcolor{red}{
    As criticality can serve as a valuable resource for quantum metrology \cite{zanardi2008,macieszczak2016,garbe2020,chu2021,Ding2022,ilias2022,Garbe_2022,Gietka2022understanding,Aybar2022criticalquantum,guan2021,wang2024,xue2026},
recent work has harnessed dynamical quantum phase transitions to achieve
high sensitivity of
% finding applications in 
parameter estimation in
% at  with 
spinor condensates \cite{zhouPRR}.
% }
In this context, critical regimes of the effective Hamiltonian provide a natural mechanism to generate squeezing and entanglement. In the following, we consider two scenarios based on ground-state and excited-state criticality.

\begin{figure}
[htb]\centering
\includegraphics[width=3 in]{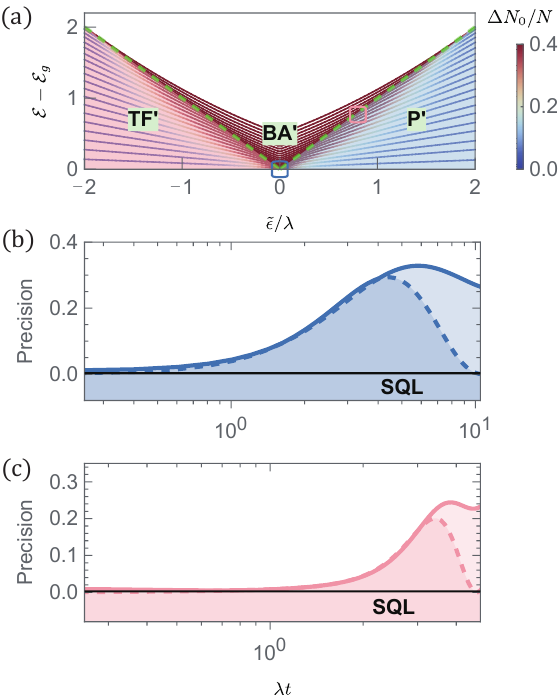}
\caption{Criticality-enhanced quantum sensing in the interferometric protocol.
(a) Excited-state phase diagram of the effective spinor Hamiltonian (\ref{eq:H_eff}) for $N = 100$ atoms at zero magnetization, shown as a function of the scaled energy $\mathcal{E}$ and the effective quadratic Zeeman parameter $\tilde{\epsilon}$. The color scale represents the normalized number variance $\Delta N_0 / N$ of the dressed mode $\varphi_0$ for the eigenstates. The green dashed lines indicate the ESQPT lines.
(b),(c) Estimation precision and quantum Fisher information for initial states prepared near the ground-state quantum phase transition (blue square in (a), panel (b)) and near an ESQPT (red square in (a), panel (c)), for $N = 100$ atoms. The dashed lines show the rescaled precision $1/[4N^2(\Delta \theta)^2]$, obtained from the error-propagation formula (\ref{eq:error-propagation}) for $\mathcal{O}=Q_{yz}$, plotted as a function of the entanglement preparation time $t$, while the solid lines show the corresponding QFI $\mathcal{I}_\theta$. The black horizontal lines indicate the standard quantum limit (SQL), $\mathcal{I}_\theta \sim N$.
}
\label{fig:qfi}
\end{figure}

A sketch of the excited-state phase diagram is shown in Fig.~\ref{fig:qfi}(a), obtained via exact diagonalization of the effective Hamiltonian (\ref{eq:H_eff}) for $N = 100$ atoms. Here $\mathcal{E} = \langle \hat{H} \rangle/(|\lambda| N)$ denotes the scaled energy per particle of the eigenstates of $\hat{H}$, and $\mathcal{E}_g$ is the corresponding ground-state energy. For illustration, we focus on the antiferromagnetic regime of the effective model ($\lambda > 0$), which can be accessed by tuning the Raman coupling even for a bare ferromagnetic gas ($g_2 < 0$). In this regime, the spectrum exhibits a dense accumulation of avoided crossings along $\mathcal{E} - \mathcal{E}_g = |\tilde{\epsilon}|$, signaling the presence of excited-state quantum phase transitions (ESQPTs). In the vicinity of these lines, the eigenstates display enhanced fluctuations, as reflected in the increased variance $\Delta N_0$, which can be harnessed for metrological purposes.

We first consider the use of ground-state criticality to generate entanglement. This can be achieved by initializing the system in the polar (P) state and tuning $\tilde{\epsilon}$ close to zero, where the transition from the polar to the twin-Fock (TF) phase occurs, as indicated by the blue square in Fig.~\ref{fig:qfi}(a). In this regime, the dynamics generate squeezing and enhance quantum fluctuations.

To ensure a non-vanishing sensitivity, we consider a rotated interferometric scheme in which the phase imprinting is performed with respect to a quadrupolar operator. This can be implemented via a rotation $R = e^{-i \pi L_y / 2} e^{-i \pi D_{xy} / 2}$, where $D_{xy} =
b_{-1}^\dagger b_1 + b_1^\dagger b_{-1}$, such that $ R^\dagger L_z R = Q_{yz}$. Here, the quadrupole operators are defined as $ Q_{ij} = b_\alpha^\dagger \left(F_i F_j + F_j F_i - \frac{4}{3} \delta_{ij} \right) b_\beta$, with repeated indices summed over the spin components. The generator $\Lambda = Q_{yz}$ has been shown to be an optimal choice for parameter estimation in spin-1 systems \cite{Niezgoda_2019}. This can be understood from the fact that, for an initial polar state, $L_x$ and $Q_{yz}$ form a pair of observables exhibiting spin-nematic squeezing \cite{Hamley2012}. Imprinting the phase along the principal squeezing axis therefore maximizes the relevant quantum fluctuations and enhances the sensitivity to parameter variations.

The corresponding QFI $\mathcal{I}_\theta = 4 (\Delta Q_{yz})^2$ is evaluated using the truncated Wigner approximation (see Appendix~\ref{app_twa} for details), and the resulting dynamics are shown in Fig.~\ref{fig:qfi}(b). The initial value of $\mathcal{I}_\theta$ exceeds the standard quantum limit (SQL) $\mathcal{I}_\theta \sim N$, indicating that the state exhibits useful quantum correlations. During the time evolution, the QFI initially increases exponentially at short times before reaching a maximum. This behavior indicates that Heisenberg-limited sensitivity can be achieved within the protocol when the evolution time is optimized alongside the atom number $N$.

In practice, the QFI is not directly measurable, and the estimation precision $\Delta \theta$ can be accessed via an observable $\mathcal{O}$ through the standard error-propagation formula
\begin{equation}
    (\Delta \theta)_\mathcal{O}^{-2}
    =
    \frac{|\partial_\theta \langle \mathcal{O} \rangle|^2}{\Delta^2 \mathcal{O}}
    \le \mathcal{I}_\theta,
\label{eq:error-propagation}
\end{equation}
where expectation values are taken with respect to the state $\vert \psi(\theta) \rangle = e^{iHt} e^{-i \theta Q_{yz}} e^{-iHt} \vert \psi_0 \rangle$, which corresponds to the echo sequence introduced above, in the limit $\theta \to 0$. For $\mathcal{O} = Q_{yz}$, the resulting precision is also shown in Fig.~\ref{fig:qfi}(b), 
alongside with $\mathcal{I}_\theta$. We plot the rescaled quantity $1/[4 N^2 (\Delta \theta)^2]$, which reaches the QFI bound at short times and exhibits Heisenberg scaling near its maximum. Although the precision rapidly degrades at longer times, its short-time behavior indicates that $Q_{yz}$ provides an effective observable for interferometric measurements. Similar observables and criticality-enhanced sensitivity have been discussed in related spinor-condensate settings \cite{zhouPRR}.

We next turn to excited-state criticality. This case is relevant because the Raman-dressed gas can retain the critical structures of its effective spinor across its excited spectrum, as discussed in our previous work \cite{Cabedo2021}. To illustrate this possibility, we consider an initial coherent spin state prepared in the vicinity of an ESQPT, as indicated by the red square in Fig.~\ref{fig:qfi}(a), at $\tilde{\epsilon}=0.6\lambda$. The state is taken as
\begin{equation}
    \vert \bm{\zeta} \rangle^{\otimes N}
    =
    \frac{
    (\zeta_1 b_1^\dagger
    +
    \zeta_0 b_0^\dagger
    +
    \zeta_{-1} b_{-1}^\dagger
    )^N
    }{
        \sqrt{N!}
    }
    \vert \text{vac} \rangle,
\end{equation}
with $\bm{\zeta} = \left(\sqrt{\frac{1-\rho_0}{2}},
\sqrt{\rho_0}, \sqrt{\frac{1-\rho_0}{2}} \right)^T$, $\rho_0 = 0.7$.
The corresponding dynamics of the QFI and estimation precision are shown in Fig.~\ref{fig:qfi}(c). The behavior is qualitatively similar to the ground-state critical case, but, remarkably, the maximum QFI and precision are reached on shorter timescales, although with slightly smaller peak values.

These results highlight that a key advantage of the present platform lies in the controllability of the spin-mixing interaction $\lambda$, including the ability to tune its magnitude and, in particular, reverse its sign. This enables the implementation of echo-type interferometric protocols based on effective time reversal. At the same time, within this framework, the effective spinor structure of the Raman-dressed gas provides access to a wide range of critical regimes, including both ground-state transitions and excited-state quantum phase transitions, which can be exploited to generate entanglement and enhance metrological sensitivity beyond the SQL, approaching the Heisenberg limit.

In the following section, we further explore how, in the Raman-dressed spin-1 Bose gas, the coupling between internal (spin) and external (momentum) degrees of freedom enables alternative phase-readout strategies based on spatial density modulations, beyond conventional spin observables.

% =============================
\section{Phase readout via SOC-induced density modulations}
\label{sec:stripe_phase}

We now turn to the measurement stage and discuss how SOC-induced spatial structures provide an alternative readout of the interferometric phase. To illustrate this point, we consider the phase imprinting generated by the operator $\Lambda = L_z$. In this case, however, the effective Hamiltonian~(\ref{eq:H_eff}) preserves an SO(2) symmetry associated with rotations around the $z$ axis, implying that $L_z$ is conserved during the dynamics. For initial states with zero magnetization, $\langle L_z \rangle = 0$, such as the polar (P) state, this results in a vanishing quantum Fisher information $\mathcal{I}_\theta = 4(\Delta L_z)^2$ within the effective spinor description. This reflects the fact that spin-based observables are insensitive to the imprinted phase under these conditions.

Importantly, this does not imply that the phase information is absent in the full system. In the SOC setting, the dressed modes $\varphi_j$ carry nontrivial quasimomentum components, and the coupling between spin and motion provides an additional channel through which the phase can be encoded and detected. In particular, relative phases between the modes $\varphi_{\pm 1}$ manifest as spatial interference patterns in the condensate density.

\begin{figure}
[htb]\centering
\includegraphics[width=3 in]{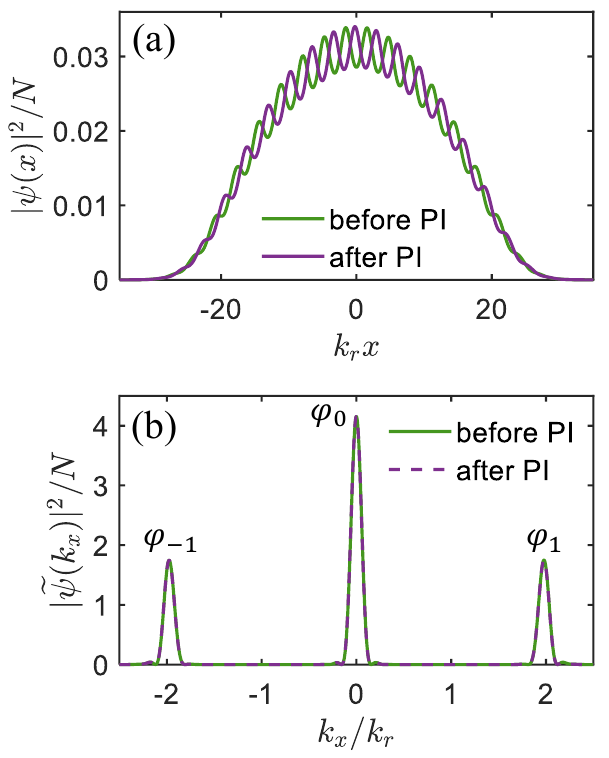} 
\caption{
Extraction of the imprinted phase from SOC-induced density modulations.
(a) Condensate density profiles and (b) corresponding quasimomentum distributions before and after phase imprinting (PI). Phase imprinting generated by $L_z$ induces a relative phase between the dressed modes, resulting in a shift of the density modulation pattern.
% \textcolor{red}{
    Before PI the parameters are set as
    $\delta = 0$,
    $\Omega = 0.65 E_r$ and
    $\epsilon = -0.03 E_r$,
    result in
    $\lambda = 0.0035 E_r$.
    During PI the parameters are quenched to
    $\delta = 0.08 E_r$ and 
    $\Omega = \Omega_c$,
    thus making $\lambda = 0$.
% }
Other parameters for the GPE simulations are specified in Appendix~\ref{app_GP}.
}
\label{fig:stripe}
\end{figure}

To illustrate this mechanism, we consider the real-space dynamics of the condensate obtained from the GPE (see Appendix~\ref{app_GP} for details). The system is initialized in the polar state of the dressed basis, corresponding to the macroscopic occupation of the mode $\varphi_0$. We first evolve the full wavefunction under the GPE for $1.1$ ms, with $\delta=0$. During this stage, the spin-mixing dynamics transfers population from $\varphi_0$ to the modes $\varphi_{\pm 1}$, leading to the formation of a spatial density modulation, shown as the green curve in Fig.~\ref{fig:stripe}(a). The corresponding quasimomentum distribution is shown in Fig.~\ref{fig:stripe}(b). This density pattern arises from the interference between components with quasimomentum $\pm 2k_r$, associated with plane-wave factors $\exp(\pm 2 i k_r x)$.

To implement phase imprinting, the parameters are then quenched to a regime with $\lambda=0$ and $\delta=0.08E_r$, and the system is evolved for $1.5$ ms. The resulting density profile is shown as the purple curve in Fig.~\ref{fig:stripe}(a). Compared with the profile before phase imprinting, the stripe pattern is shifted, while the quasimomentum distribution in Fig.~\ref{fig:stripe}(b) remains essentially unchanged. This is consistent with the fact that the phase-imprinting stage mainly changes the relative phase of the dressed modes $\varphi_{\pm1}$, which acquire opposite phase factors $\exp(\pm i\theta)$, without significantly changing their populations. The imprinted phase $\theta$ can therefore be extracted by comparing the stripe position before and after phase imprinting.

Remarkably, the vanishing QFI discussed above reflects the insensitivity of spin observables within the effective spinor model, rather than the absence of phase information in the full system. The effective Hamiltonian~(\ref{eq:H_eff}) is obtained by integrating out the spatial degrees of freedom associated with the dressed states, and therefore does not capture observables that depend explicitly on the condensate density profile. In this sense, SOC provides access to a complementary measurement channel that exploits the coupling between internal and external degrees of freedom.

While the present mean-field simulations demonstrate phase extraction at the level of density profiles, the underlying momentum-space correlations generated by the interacting dynamics suggest that, in principle, spatial observables could access quantum-enhanced sensitivity. A quantitative assessment of such effects would require going beyond the mean-field description, for instance by evaluating density correlations derived from the full quantum state, and is left for future work.
% =============================

% summary   ===================
\section{Discussion and outlook}
\label{sec:conclusion}

In summary, we have proposed quantum interferometric protocols based on a Raman-dressed spin-1 Bose gas with spin--orbit coupling. In the low-energy regime, this system realizes an effective spinor Hamiltonian with a spin-mixing interaction $\lambda$ that can be controlled through the Raman coupling. This tunability provides additional flexibility compared with conventional spinor condensates: it allows the spin-mixing dynamics to be suppressed during phase imprinting and may enable the sign reversal of $\lambda$, opening a route to echo-type protocols based on effective time reversal.

We have analyzed how these protocols can benefit from critical regimes of the effective spinor Hamiltonian. In particular, ground-state and excited-state quantum critical regimes can generate large fluctuations and entanglement, leading to metrological gains beyond the standard quantum limit and approaching the Heisenberg limit within the echo protocol. These results show that the effective spinor structure realized in the Raman-dressed gas provides a useful framework for nonlinear atomic interferometry.

A second distinctive feature of the platform is that the dressed modes retain the spin--momentum structure inherited from spin--orbit coupling. As a consequence, the interferometric phase can be encoded not only in collective spin observables, but also in spatial density modulations. We illustrated this mechanism with Gross--Pitaevskii simulations showing that phase imprinting can shift the density stripe pattern, thus providing an alternative real-space readout of the phase. This spatial readout should be understood as complementary to the spin-based protocols. While the mean-field simulations demonstrate phase extraction at the level of density profiles, a quantitative assessment of quantum-enhanced sensitivity through spatial observables would require going beyond mean-field theory.

Several directions remain open. A particularly important one is to develop a unified quantum-metrological treatment that includes both the effective spinor dynamics and the spatial degrees of freedom of the Raman-dressed gas. Such an approach would make it possible to quantify whether spatial observables, such as density correlations associated with the stripe pattern, can access quantum-enhanced sensitivity beyond the mean-field level. This would further clarify the role of spin--momentum locking as a resource for quantum sensing.
It will also be interesting to explore quantum metrology with spin-momentum correlated atoms in a cavity setup \cite{Oksana2024,Finger2024},
in which criticality also play a role.

% =============================

\begin{acknowledgments}
% We thank Han Pu for helpful discussions. 
This work is supported by the
Quantum Science and Technology-National Science and Technology Major Project (2021ZD0303200);
the National Natural Science Foundation of China (Grant Nos.
12074120,
%11374003,
%11654005,
12374328,  12234014, 12005049, 11935012), the Shanghai Municipal Science and
Technology Major Project (Grant No. 2019SHZDZX01), Innovation Program of
Shanghai Municipal Education Commission (Grant No. 202101070008E00099),
Shanghai Science and Technology Innovation Project (No. 24LZ1400600), and the
Fundamental Research Funds for the Central Universities,
National Key Research and Development Program of China (Grant No.
2016YFA0302001). 
J.C. acknowledges support from the
Ministry of Economic Affairs and Digital Transformation of the Spanish Government through
the QUANTUM ENIA project call – Quantum Spain project. A.C. acknowledges
funding from the Spanish Ministry of Science and Innovation
MCIN/AEI/10.13039/501100011033 (project MAPS PID2023-149988NB-C21), the EU
QuantERA project DYNAMITE (funded by MICN/AEI/ 10.13039/501100011033 and by
the European Union NextGenerationEU/PRTR PCI2022-132919 (Grant No.
101017733)), and the Generalitat de Catalunya (AGAUR SGR 2021- SGR-00138)).
W.Z. acknowledges additional support from the Shanghai Talent Program.
%National Natural Science
%Foundation of China (Grants No. , No. 11774093) and the National Key
%Research and Development Program of China (Grant No. 2016YFA0302001).
L.Z. acknowledges additional support from China Scholarship Council.
%the Natural Science Foundation of Shanghai (Grant No. 20ZR1418500).

\end{acknowledgments}

\appendix

% twa   =======================
\section{Truncated Wigner Approximation (TWA)}
\label{app_twa}

% blablabla
We adopt TWA to
study the dynamics and obtain the results presented in Fig.~\ref{fig:qfi}.
TWA states that the Wigner function $W$ for a
quantum state approximately follows the equation%
\begin{equation}
i\hbar\frac{\partial W}{\partial t}\simeq\left\{ H_{W},W\right\} _{C},
\label{eq:twa1}
\end{equation}
where $H_{W}$ is the Wigner-Weyl transform of the Hamiltonian (\ref{eq:H_eff}), and $\left\{
\cdots\right\} _{C}$ is the coherent state Poisson bracket. 
% Similarly 
In the
coherent state picture we treat the operators $b_{j}$ ($b%
_{j}^{\dagger}$) as complex $c$-numbers $\beta_{j}$ ($\beta_{j}^{\ast}$),
and making Wigner-Weyl transform to the Heisenberg equations, 
we have

\begin{equation}
i\hbar \frac{\partial \beta _{j}}{\partial t}\simeq \left\{ \beta
_{j},H_{W}\right\} _{C}=\frac{\partial H_{W}}{\partial \beta _{j}^{\ast }}.
\label{eq:twa2}
\end{equation}%
TWA then involves first sampling the Wigner distribution $W$ with many sets
of $\left\{ \beta _{j},\beta _{j}^{\ast }\right\} $, and then for each set
we solve the equation of motion (\ref{eq:twa2}). 

In practice we sample a system of $N=100$ with $500$ trajectories. It has been compared
with the exact quantum mechanical calculations regarding the QFI, where good agreements are
found in short time scale. 
% =============================

% gp solver ===================
\section{Solve GPE }
\label{app_GP}

% \textcolor{blue}{give details on how to solve the gp equation,
% may give some figures}
We simulate the GPEs for the whole system:
\begin{equation}
    i\hbar\dot{\phi}_{j}=\partial\mathcal{E}/\partial\phi_{j}^{\ast}
\label{eq:GPE}
\end{equation} 
with $\mathcal{E} = \boldsymbol{\psi} (\mathcal{H}_{\text{SOC}} + V_t) \boldsymbol{\psi} + \frac{g_0}{2} (\boldsymbol{\psi}^\dagger \boldsymbol{\psi})^2 + \frac{g_2}{2} \sum_j (\boldsymbol{\psi} F_j \boldsymbol{\psi})^2$.
We consider that the gas is confined within an
isotropic harmonic potential $V_t=\frac{1}{2}m\omega_{t}^{2}\boldsymbol{r}^{2}$
with $\omega_{t} = 2\pi\times140$ Hz. 
For $^{87}\text{Rb}$ at
$n = 7.5 \times 10^{13} \, \text{cm}^{-3}$,
$g_0 k_r^3 = 1.066 \, E_r$ and
$g_2 / g_0 = -0.0047$,
in which
$k_r = 7.95 \times 10^6 \text{m}^{-1}$,
$E_r / \hbar = 2 \pi \times 3680 \, \text{Hz}$.
Other parameters are set as $N=10^{4}$ and
$\epsilon = -0.03 \, E_r$.
 % and 
 % and $E_{r}/\hbar=2\pi\times3680$ Hz.

In numerical simulation, 
we adopt the split-step Crank-Nicolson scheme \cite{RAVISANKAR2021107657}.
First the dressed states $\bm{\varphi}$ are derived via imaginary time evolution of the GPE.
Then the dynamics presented in Fig.~\ref{fig:stripe} are obtained via propagating the GPE in real time with an initial state
$\bm{\psi} = \varphi_0$.
In the entanglement preparation process we set $\Omega=0.65E_{r}$,
then $\Omega$ is quenched to the value of $\Omega_c$ to ensure the effective spin-mixing interaction strength $\lambda = 0$ during phase imprinting.
% =============================

% \bibliographystyle{plain}
% \bibliography{bib-moire}

\begin{thebibliography}{54}%
\makeatletter
\providecommand \@ifxundefined [1]{%
 \@ifx{#1\undefined}
}%
\providecommand \@ifnum [1]{%
 \ifnum #1\expandafter \@firstoftwo
 \else \expandafter \@secondoftwo
 \fi
}%
\providecommand \@ifx [1]{%
 \ifx #1\expandafter \@firstoftwo
 \else \expandafter \@secondoftwo
 \fi
}%
\providecommand \natexlab [1]{#1}%
\providecommand \enquote  [1]{``#1''}%
\providecommand \bibnamefont  [1]{#1}%
\providecommand \bibfnamefont [1]{#1}%
\providecommand \citenamefont [1]{#1}%
\providecommand \href@noop [0]{\@secondoftwo}%
\providecommand \href [0]{\begingroup \@sanitize@url \@href}%
\providecommand \@href[1]{\@@startlink{#1}\@@href}%
\providecommand \@@href[1]{\endgroup#1\@@endlink}%
\providecommand \@sanitize@url [0]{\catcode `\\12\catcode `\$12\catcode
  `\&12\catcode `\#12\catcode `\^12\catcode `\_12\catcode `\%12\relax}%
\providecommand \@@startlink[1]{}%
\providecommand \@@endlink[0]{}%
\providecommand \url  [0]{\begingroup\@sanitize@url \@url }%
\providecommand \@url [1]{\endgroup\@href {#1}{\urlprefix }}%
\providecommand \urlprefix  [0]{URL }%
\providecommand \Eprint [0]{\href }%
\providecommand \doibase [0]{http://dx.doi.org/}%
\providecommand \selectlanguage [0]{\@gobble}%
\providecommand \bibinfo  [0]{\@secondoftwo}%
\providecommand \bibfield  [0]{\@secondoftwo}%
\providecommand \translation [1]{[#1]}%
\providecommand \BibitemOpen [0]{}%
\providecommand \bibitemStop [0]{}%
\providecommand \bibitemNoStop [0]{.\EOS\space}%
\providecommand \EOS [0]{\spacefactor3000\relax}%
\providecommand \BibitemShut  [1]{\csname bibitem#1\endcsname}%
\let\auto@bib@innerbib\@empty
%</preamble>
\bibitem [{\citenamefont {Stamper-Kurn}\ and\ \citenamefont
  {Ueda}(2013)}]{Stamper-Kurn2013}%
  \BibitemOpen
  \bibfield  {author} {\bibinfo {author} {\bibfnamefont {D.~M.}\ \bibnamefont
  {Stamper-Kurn}}\ and\ \bibinfo {author} {\bibfnamefont {M.}~\bibnamefont
  {Ueda}},\ }\bibfield  {title} {\bibinfo {title} {Spinor bose gases:
  Symmetries, magnetism, and quantum dynamics},\ }\href {\doibase
  10.1103/RevModPhys.85.1191} {\bibfield  {journal} {\bibinfo  {journal} {Rev.
  Mod. Phys.}\ }\textbf {\bibinfo {volume} {85}},\ \bibinfo {pages} {1191}
  (\bibinfo {year} {2013})}\BibitemShut {NoStop}%
\bibitem [{\citenamefont {Pu}\ and\ \citenamefont {Meystre}(2000)}]{Pu2000}%
  \BibitemOpen
  \bibfield  {author} {\bibinfo {author} {\bibfnamefont {H.}~\bibnamefont
  {Pu}}\ and\ \bibinfo {author} {\bibfnamefont {P.}~\bibnamefont {Meystre}},\
  }\bibfield  {title} {\bibinfo {title} {Creating macroscopic atomic
  einstein-podolsky-rosen states from bose-einstein condensates},\ }\href
  {\doibase 10.1103/PhysRevLett.85.3987} {\bibfield  {journal} {\bibinfo
  {journal} {Phys. Rev. Lett.}\ }\textbf {\bibinfo {volume} {85}},\ \bibinfo
  {pages} {3987} (\bibinfo {year} {2000})}\BibitemShut {NoStop}%
\bibitem [{\citenamefont {Duan}\ \emph {et~al.}(2000)\citenamefont {Duan},
  \citenamefont {S\o{}rensen}, \citenamefont {Cirac},\ and\ \citenamefont
  {Zoller}}]{Duan2000}%
  \BibitemOpen
  \bibfield  {author} {\bibinfo {author} {\bibfnamefont {L.-M.}\ \bibnamefont
  {Duan}}, \bibinfo {author} {\bibfnamefont {A.}~\bibnamefont {S\o{}rensen}},
  \bibinfo {author} {\bibfnamefont {J.~I.}\ \bibnamefont {Cirac}}, \ and\
  \bibinfo {author} {\bibfnamefont {P.}~\bibnamefont {Zoller}},\ }\bibfield
  {title} {\bibinfo {title} {Squeezing and entanglement of atomic beams},\
  }\href {\doibase 10.1103/PhysRevLett.85.3991} {\bibfield  {journal} {\bibinfo
   {journal} {Phys. Rev. Lett.}\ }\textbf {\bibinfo {volume} {85}},\ \bibinfo
  {pages} {3991} (\bibinfo {year} {2000})}\BibitemShut {NoStop}%
\bibitem [{\citenamefont {Hamley}\ \emph {et~al.}(2012)\citenamefont {Hamley},
  \citenamefont {Gerving}, \citenamefont {Hoang}, \citenamefont {Bookjans},\
  and\ \citenamefont {Chapman}}]{Hamley2012}%
  \BibitemOpen
  \bibfield  {author} {\bibinfo {author} {\bibfnamefont {C.~D.}\ \bibnamefont
  {Hamley}}, \bibinfo {author} {\bibfnamefont {C.~S.}\ \bibnamefont {Gerving}},
  \bibinfo {author} {\bibfnamefont {T.~M.}\ \bibnamefont {Hoang}}, \bibinfo
  {author} {\bibfnamefont {E.~M.}\ \bibnamefont {Bookjans}}, \ and\ \bibinfo
  {author} {\bibfnamefont {M.~S.}\ \bibnamefont {Chapman}},\ }\bibfield
  {title} {\bibinfo {title} {Spin-nematic squeezed vacuum in a quantum gas},\
  }\href {\doibase 10.1038/nphys2245} {\bibfield  {journal} {\bibinfo
  {journal} {Nat. Phys.}\ }\textbf {\bibinfo {volume} {8}},\ \bibinfo {pages}
  {305} (\bibinfo {year} {2012})}\BibitemShut {NoStop}%
\bibitem [{\citenamefont {Luo}\ \emph {et~al.}(2017)\citenamefont {Luo},
  \citenamefont {Zou}, \citenamefont {Wu}, \citenamefont {Liu}, \citenamefont
  {Han}, \citenamefont {Tey},\ and\ \citenamefont {You}}]{Luo2017}%
  \BibitemOpen
  \bibfield  {author} {\bibinfo {author} {\bibfnamefont {X.-Y.}\ \bibnamefont
  {Luo}}, \bibinfo {author} {\bibfnamefont {Y.-Q.}\ \bibnamefont {Zou}},
  \bibinfo {author} {\bibfnamefont {L.-N.}\ \bibnamefont {Wu}}, \bibinfo
  {author} {\bibfnamefont {Q.}~\bibnamefont {Liu}}, \bibinfo {author}
  {\bibfnamefont {M.-F.}\ \bibnamefont {Han}}, \bibinfo {author} {\bibfnamefont
  {M.~K.}\ \bibnamefont {Tey}}, \ and\ \bibinfo {author} {\bibfnamefont
  {L.}~\bibnamefont {You}},\ }\bibfield  {title} {\bibinfo {title}
  {Deterministic entanglement generation from driving through quantum phase
  transitions},\ }\href {\doibase 10.1126/science.aag1106} {\bibfield
  {journal} {\bibinfo  {journal} {Science}\ }\textbf {\bibinfo {volume}
  {355}},\ \bibinfo {pages} {620} (\bibinfo {year} {2017})}\BibitemShut
  {NoStop}%
\bibitem [{\citenamefont {Hudelist}\ \emph {et~al.}(2014)\citenamefont
  {Hudelist}, \citenamefont {Kong}, \citenamefont {Liu}, \citenamefont {Jing},
  \citenamefont {Ou},\ and\ \citenamefont {Zhang}}]{Hudelist2014}%
  \BibitemOpen
  \bibfield  {author} {\bibinfo {author} {\bibfnamefont {F.}~\bibnamefont
  {Hudelist}}, \bibinfo {author} {\bibfnamefont {J.}~\bibnamefont {Kong}},
  \bibinfo {author} {\bibfnamefont {C.}~\bibnamefont {Liu}}, \bibinfo {author}
  {\bibfnamefont {J.}~\bibnamefont {Jing}}, \bibinfo {author} {\bibfnamefont
  {Z.~Y.}\ \bibnamefont {Ou}}, \ and\ \bibinfo {author} {\bibfnamefont
  {W.}~\bibnamefont {Zhang}},\ }\bibfield  {title} {\bibinfo {title} {Quantum
  metrology with parametric amplifier-based photon correlation
  interferometers},\ }\href {\doibase 10.1038/ncomms4049} {\bibfield  {journal}
  {\bibinfo  {journal} {Nat. Commun.}\ }\textbf {\bibinfo {volume} {5}},\
  \bibinfo {pages} {3049} (\bibinfo {year} {2014})}\BibitemShut {NoStop}%
\bibitem [{\citenamefont {Gabbrielli}\ \emph {et~al.}(2015)\citenamefont
  {Gabbrielli}, \citenamefont {Pezz\`e},\ and\ \citenamefont
  {Smerzi}}]{Gabbrielli2015}%
  \BibitemOpen
  \bibfield  {author} {\bibinfo {author} {\bibfnamefont {M.}~\bibnamefont
  {Gabbrielli}}, \bibinfo {author} {\bibfnamefont {L.}~\bibnamefont {Pezz\`e}},
  \ and\ \bibinfo {author} {\bibfnamefont {A.}~\bibnamefont {Smerzi}},\
  }\bibfield  {title} {\bibinfo {title} {Spin-mixing interferometry with
  bose-einstein condensates},\ }\href {\doibase 10.1103/PhysRevLett.115.163002}
  {\bibfield  {journal} {\bibinfo  {journal} {Phys. Rev. Lett.}\ }\textbf
  {\bibinfo {volume} {115}},\ \bibinfo {pages} {163002} (\bibinfo {year}
  {2015})}\BibitemShut {NoStop}%
\bibitem [{\citenamefont {Linnemann}\ \emph {et~al.}(2016)\citenamefont
  {Linnemann}, \citenamefont {Strobel}, \citenamefont {Muessel}, \citenamefont
  {Schulz}, \citenamefont {Lewis-Swan}, \citenamefont {Kheruntsyan},\ and\
  \citenamefont {Oberthaler}}]{Linnemann2016}%
  \BibitemOpen
  \bibfield  {author} {\bibinfo {author} {\bibfnamefont {D.}~\bibnamefont
  {Linnemann}}, \bibinfo {author} {\bibfnamefont {H.}~\bibnamefont {Strobel}},
  \bibinfo {author} {\bibfnamefont {W.}~\bibnamefont {Muessel}}, \bibinfo
  {author} {\bibfnamefont {J.}~\bibnamefont {Schulz}}, \bibinfo {author}
  {\bibfnamefont {R.~J.}\ \bibnamefont {Lewis-Swan}}, \bibinfo {author}
  {\bibfnamefont {K.~V.}\ \bibnamefont {Kheruntsyan}}, \ and\ \bibinfo {author}
  {\bibfnamefont {M.~K.}\ \bibnamefont {Oberthaler}},\ }\bibfield  {title}
  {\bibinfo {title} {Quantum-enhanced sensing based on time reversal of
  nonlinear dynamics},\ }\href {\doibase 10.1103/PhysRevLett.117.013001}
  {\bibfield  {journal} {\bibinfo  {journal} {Phys. Rev. Lett.}\ }\textbf
  {\bibinfo {volume} {117}},\ \bibinfo {pages} {013001} (\bibinfo {year}
  {2016})}\BibitemShut {NoStop}%
\bibitem [{\citenamefont {Wrubel}\ \emph {et~al.}(2018)\citenamefont {Wrubel},
  \citenamefont {Schwettmann}, \citenamefont {Fahey}, \citenamefont {Glassman},
  \citenamefont {Pechkis}, \citenamefont {Griffin}, \citenamefont {Barnett},
  \citenamefont {Tiesinga},\ and\ \citenamefont {Lett}}]{Wrubel2018}%
  \BibitemOpen
  \bibfield  {author} {\bibinfo {author} {\bibfnamefont {J.~P.}\ \bibnamefont
  {Wrubel}}, \bibinfo {author} {\bibfnamefont {A.}~\bibnamefont {Schwettmann}},
  \bibinfo {author} {\bibfnamefont {D.~P.}\ \bibnamefont {Fahey}}, \bibinfo
  {author} {\bibfnamefont {Z.}~\bibnamefont {Glassman}}, \bibinfo {author}
  {\bibfnamefont {H.~K.}\ \bibnamefont {Pechkis}}, \bibinfo {author}
  {\bibfnamefont {P.~F.}\ \bibnamefont {Griffin}}, \bibinfo {author}
  {\bibfnamefont {R.}~\bibnamefont {Barnett}}, \bibinfo {author} {\bibfnamefont
  {E.}~\bibnamefont {Tiesinga}}, \ and\ \bibinfo {author} {\bibfnamefont
  {P.~D.}\ \bibnamefont {Lett}},\ }\bibfield  {title} {\bibinfo {title} {Spinor
  bose-einstein-condensate phase-sensitive amplifier for su(1,1)
  interferometry},\ }\href {\doibase 10.1103/PhysRevA.98.023620} {\bibfield
  {journal} {\bibinfo  {journal} {Phys. Rev. A}\ }\textbf {\bibinfo {volume}
  {98}},\ \bibinfo {pages} {023620} (\bibinfo {year} {2018})}\BibitemShut
  {NoStop}%
\bibitem [{\citenamefont {Qu}\ \emph {et~al.}(2020)\citenamefont {Qu},
  \citenamefont {Evrard}, \citenamefont {Dalibard},\ and\ \citenamefont
  {Gerbier}}]{Qu2020}%
  \BibitemOpen
  \bibfield  {author} {\bibinfo {author} {\bibfnamefont {A.}~\bibnamefont
  {Qu}}, \bibinfo {author} {\bibfnamefont {B.}~\bibnamefont {Evrard}}, \bibinfo
  {author} {\bibfnamefont {J.}~\bibnamefont {Dalibard}}, \ and\ \bibinfo
  {author} {\bibfnamefont {F.}~\bibnamefont {Gerbier}},\ }\bibfield  {title}
  {\bibinfo {title} {Probing spin correlations in a bose-einstein condensate
  near the single-atom level},\ }\href {\doibase
  10.1103/PhysRevLett.125.033401} {\bibfield  {journal} {\bibinfo  {journal}
  {Phys. Rev. Lett.}\ }\textbf {\bibinfo {volume} {125}},\ \bibinfo {pages}
  {033401} (\bibinfo {year} {2020})}\BibitemShut {NoStop}%
\bibitem [{\citenamefont {Law}\ \emph {et~al.}(1998)\citenamefont {Law},
  \citenamefont {Pu},\ and\ \citenamefont {Bigelow}}]{Law1998}%
  \BibitemOpen
  \bibfield  {author} {\bibinfo {author} {\bibfnamefont {C.~K.}\ \bibnamefont
  {Law}}, \bibinfo {author} {\bibfnamefont {H.}~\bibnamefont {Pu}}, \ and\
  \bibinfo {author} {\bibfnamefont {N.~P.}\ \bibnamefont {Bigelow}},\
  }\bibfield  {title} {\bibinfo {title} {Quantum spins mixing in spinor
  bose-einstein condensates},\ }\href {\doibase 10.1103/PhysRevLett.81.5257}
  {\bibfield  {journal} {\bibinfo  {journal} {Phys. Rev. Lett.}\ }\textbf
  {\bibinfo {volume} {81}},\ \bibinfo {pages} {5257} (\bibinfo {year}
  {1998})}\BibitemShut {NoStop}%
\bibitem [{\citenamefont {Liu}\ \emph {et~al.}(2022)\citenamefont {Liu},
  \citenamefont {Wu}, \citenamefont {Cao}, \citenamefont {Mao}, \citenamefont
  {Li}, \citenamefont {Guo}, \citenamefont {Tey},\ and\ \citenamefont
  {You}}]{Liu2022}%
  \BibitemOpen
  \bibfield  {author} {\bibinfo {author} {\bibfnamefont {Q.}~\bibnamefont
  {Liu}}, \bibinfo {author} {\bibfnamefont {L.-N.}\ \bibnamefont {Wu}},
  \bibinfo {author} {\bibfnamefont {J.-H.}\ \bibnamefont {Cao}}, \bibinfo
  {author} {\bibfnamefont {T.-W.}\ \bibnamefont {Mao}}, \bibinfo {author}
  {\bibfnamefont {X.-W.}\ \bibnamefont {Li}}, \bibinfo {author} {\bibfnamefont
  {S.-F.}\ \bibnamefont {Guo}}, \bibinfo {author} {\bibfnamefont {M.~K.}\
  \bibnamefont {Tey}}, \ and\ \bibinfo {author} {\bibfnamefont
  {L.}~\bibnamefont {You}},\ }\bibfield  {title} {\bibinfo {title} {Nonlinear
  interferometry beyond classical limit enabled by cyclic dynamics},\ }\href
  {\doibase 10.1038/s41567-021-01441-7} {\bibfield  {journal} {\bibinfo
  {journal} {Nat. Phys.}\ }\textbf {\bibinfo {volume} {18}},\ \bibinfo {pages}
  {167} (\bibinfo {year} {2022})}\BibitemShut {NoStop}%
\bibitem [{\citenamefont {Zanardi}\ \emph {et~al.}(2008)\citenamefont
  {Zanardi}, \citenamefont {Paris},\ and\ \citenamefont
  {Campos~Venuti}}]{zanardi2008}%
  \BibitemOpen
  \bibfield  {author} {\bibinfo {author} {\bibfnamefont {P.}~\bibnamefont
  {Zanardi}}, \bibinfo {author} {\bibfnamefont {M.~G.~A.}\ \bibnamefont
  {Paris}}, \ and\ \bibinfo {author} {\bibfnamefont {L.}~\bibnamefont
  {Campos~Venuti}},\ }\bibfield  {title} {\bibinfo {title} {Quantum criticality
  as a resource for quantum estimation},\ }\href {\doibase
  10.1103/PhysRevA.78.042105} {\bibfield  {journal} {\bibinfo  {journal} {Phys.
  Rev. A}\ }\textbf {\bibinfo {volume} {78}},\ \bibinfo {pages} {042105}
  (\bibinfo {year} {2008})}\BibitemShut {NoStop}%
\bibitem [{\citenamefont {Macieszczak}\ \emph {et~al.}(2016)\citenamefont
  {Macieszczak}, \citenamefont {Gu\ifmmode \mbox{\c{t}}\else
  \c{t}\fi{}\ifmmode~\u{a}\else \u{a}\fi{}}, \citenamefont {Lesanovsky},\ and\
  \citenamefont {Garrahan}}]{macieszczak2016}%
  \BibitemOpen
  \bibfield  {author} {\bibinfo {author} {\bibfnamefont {K.}~\bibnamefont
  {Macieszczak}}, \bibinfo {author} {\bibfnamefont {M.}~\bibnamefont
  {Gu\ifmmode \mbox{\c{t}}\else \c{t}\fi{}\ifmmode~\u{a}\else \u{a}\fi{}}},
  \bibinfo {author} {\bibfnamefont {I.}~\bibnamefont {Lesanovsky}}, \ and\
  \bibinfo {author} {\bibfnamefont {J.~P.}\ \bibnamefont {Garrahan}},\
  }\bibfield  {title} {\bibinfo {title} {Dynamical phase transitions as a
  resource for quantum enhanced metrology},\ }\href {\doibase
  10.1103/PhysRevA.93.022103} {\bibfield  {journal} {\bibinfo  {journal} {Phys.
  Rev. A}\ }\textbf {\bibinfo {volume} {93}},\ \bibinfo {pages} {022103}
  (\bibinfo {year} {2016})}\BibitemShut {NoStop}%
\bibitem [{\citenamefont {Garbe}\ \emph {et~al.}(2020)\citenamefont {Garbe},
  \citenamefont {Bina}, \citenamefont {Keller}, \citenamefont {Paris},\ and\
  \citenamefont {Felicetti}}]{garbe2020}%
  \BibitemOpen
  \bibfield  {author} {\bibinfo {author} {\bibfnamefont {L.}~\bibnamefont
  {Garbe}}, \bibinfo {author} {\bibfnamefont {M.}~\bibnamefont {Bina}},
  \bibinfo {author} {\bibfnamefont {A.}~\bibnamefont {Keller}}, \bibinfo
  {author} {\bibfnamefont {M.~G.~A.}\ \bibnamefont {Paris}}, \ and\ \bibinfo
  {author} {\bibfnamefont {S.}~\bibnamefont {Felicetti}},\ }\bibfield  {title}
  {\bibinfo {title} {Critical quantum metrology with a finite-component quantum
  phase transition},\ }\href {\doibase 10.1103/PhysRevLett.124.120504}
  {\bibfield  {journal} {\bibinfo  {journal} {Phys. Rev. Lett.}\ }\textbf
  {\bibinfo {volume} {124}},\ \bibinfo {pages} {120504} (\bibinfo {year}
  {2020})}\BibitemShut {NoStop}%
\bibitem [{\citenamefont {Chu}\ \emph {et~al.}(2021)\citenamefont {Chu},
  \citenamefont {Zhang}, \citenamefont {Yu},\ and\ \citenamefont
  {Cai}}]{chu2021}%
  \BibitemOpen
  \bibfield  {author} {\bibinfo {author} {\bibfnamefont {Y.}~\bibnamefont
  {Chu}}, \bibinfo {author} {\bibfnamefont {S.}~\bibnamefont {Zhang}}, \bibinfo
  {author} {\bibfnamefont {B.}~\bibnamefont {Yu}}, \ and\ \bibinfo {author}
  {\bibfnamefont {J.}~\bibnamefont {Cai}},\ }\bibfield  {title} {\bibinfo
  {title} {Dynamic framework for criticality-enhanced quantum sensing},\ }\href
  {\doibase 10.1103/PhysRevLett.126.010502} {\bibfield  {journal} {\bibinfo
  {journal} {Phys. Rev. Lett.}\ }\textbf {\bibinfo {volume} {126}},\ \bibinfo
  {pages} {010502} (\bibinfo {year} {2021})}\BibitemShut {NoStop}%
\bibitem [{\citenamefont {Ding}\ \emph {et~al.}(2022)\citenamefont {Ding},
  \citenamefont {Liu}, \citenamefont {Shi}, \citenamefont {Guo}, \citenamefont
  {M{\o}lmer},\ and\ \citenamefont {Adams}}]{Ding2022}%
  \BibitemOpen
  \bibfield  {author} {\bibinfo {author} {\bibfnamefont {D.-S.}\ \bibnamefont
  {Ding}}, \bibinfo {author} {\bibfnamefont {Z.-K.}\ \bibnamefont {Liu}},
  \bibinfo {author} {\bibfnamefont {B.-S.}\ \bibnamefont {Shi}}, \bibinfo
  {author} {\bibfnamefont {G.-C.}\ \bibnamefont {Guo}}, \bibinfo {author}
  {\bibfnamefont {K.}~\bibnamefont {M{\o}lmer}}, \ and\ \bibinfo {author}
  {\bibfnamefont {C.~S.}\ \bibnamefont {Adams}},\ }\bibfield  {title} {\bibinfo
  {title} {Enhanced metrology at the critical point of a many-body rydberg
  atomic system},\ }\href {\doibase 10.1038/s41567-022-01777-8} {\bibfield
  {journal} {\bibinfo  {journal} {Nat. Phys.}\ }\textbf {\bibinfo {volume}
  {18}},\ \bibinfo {pages} {1447} (\bibinfo {year} {2022})}\BibitemShut
  {NoStop}%
\bibitem [{\citenamefont {Ilias}\ \emph {et~al.}(2022)\citenamefont {Ilias},
  \citenamefont {Yang}, \citenamefont {Huelga},\ and\ \citenamefont
  {Plenio}}]{ilias2022}%
  \BibitemOpen
  \bibfield  {author} {\bibinfo {author} {\bibfnamefont {T.}~\bibnamefont
  {Ilias}}, \bibinfo {author} {\bibfnamefont {D.}~\bibnamefont {Yang}},
  \bibinfo {author} {\bibfnamefont {S.~F.}\ \bibnamefont {Huelga}}, \ and\
  \bibinfo {author} {\bibfnamefont {M.~B.}\ \bibnamefont {Plenio}},\ }\bibfield
   {title} {\bibinfo {title} {Criticality-enhanced quantum sensing via
  continuous measurement},\ }\href {\doibase 10.1103/PRXQuantum.3.010354}
  {\bibfield  {journal} {\bibinfo  {journal} {PRX Quantum}\ }\textbf {\bibinfo
  {volume} {3}},\ \bibinfo {pages} {010354} (\bibinfo {year}
  {2022})}\BibitemShut {NoStop}%
\bibitem [{\citenamefont {Garbe}\ \emph {et~al.}(2022)\citenamefont {Garbe},
  \citenamefont {Abah}, \citenamefont {Felicetti},\ and\ \citenamefont
  {Puebla}}]{Garbe_2022}%
  \BibitemOpen
  \bibfield  {author} {\bibinfo {author} {\bibfnamefont {L.}~\bibnamefont
  {Garbe}}, \bibinfo {author} {\bibfnamefont {O.}~\bibnamefont {Abah}},
  \bibinfo {author} {\bibfnamefont {S.}~\bibnamefont {Felicetti}}, \ and\
  \bibinfo {author} {\bibfnamefont {R.}~\bibnamefont {Puebla}},\ }\bibfield
  {title} {\bibinfo {title} {Critical quantum metrology with fully-connected
  models: from heisenberg to kibble–zurek scaling},\ }\href {\doibase
  10.1088/2058-9565/ac6ca5} {\bibfield  {journal} {\bibinfo  {journal} {Quantum
  Sci. Technol.}\ }\textbf {\bibinfo {volume} {7}},\ \bibinfo {pages} {035010}
  (\bibinfo {year} {2022})}\BibitemShut {NoStop}%
\bibitem [{\citenamefont {Gietka}\ \emph {et~al.}(2022)\citenamefont {Gietka},
  \citenamefont {Ruks},\ and\ \citenamefont {Busch}}]{Gietka2022understanding}%
  \BibitemOpen
  \bibfield  {author} {\bibinfo {author} {\bibfnamefont {K.}~\bibnamefont
  {Gietka}}, \bibinfo {author} {\bibfnamefont {L.}~\bibnamefont {Ruks}}, \ and\
  \bibinfo {author} {\bibfnamefont {T.}~\bibnamefont {Busch}},\ }\bibfield
  {title} {\bibinfo {title} {Understanding and {I}mproving {C}ritical
  {M}etrology. {Q}uenching {S}uperradiant {L}ight-{M}atter {S}ystems {B}eyond
  the {C}ritical {P}oint},\ }\href {\doibase 10.22331/q-2022-04-27-700}
  {\bibfield  {journal} {\bibinfo  {journal} {{Quantum}}\ }\textbf {\bibinfo
  {volume} {6}},\ \bibinfo {pages} {700} (\bibinfo {year} {2022})}\BibitemShut
  {NoStop}%
\bibitem [{\citenamefont {Aybar}\ \emph {et~al.}(2022)\citenamefont {Aybar},
  \citenamefont {Niezgoda}, \citenamefont {Mirkhalaf}, \citenamefont
  {Mitchell}, \citenamefont {Benedicto~Orenes},\ and\ \citenamefont
  {Witkowska}}]{Aybar2022criticalquantum}%
  \BibitemOpen
  \bibfield  {author} {\bibinfo {author} {\bibfnamefont {E.}~\bibnamefont
  {Aybar}}, \bibinfo {author} {\bibfnamefont {A.}~\bibnamefont {Niezgoda}},
  \bibinfo {author} {\bibfnamefont {S.~S.}\ \bibnamefont {Mirkhalaf}}, \bibinfo
  {author} {\bibfnamefont {M.~W.}\ \bibnamefont {Mitchell}}, \bibinfo {author}
  {\bibfnamefont {D.}~\bibnamefont {Benedicto~Orenes}}, \ and\ \bibinfo
  {author} {\bibfnamefont {E.}~\bibnamefont {Witkowska}},\ }\bibfield  {title}
  {\bibinfo {title} {Critical quantum thermometry and its feasibility in spin
  systems},\ }\href {\doibase 10.22331/q-2022-09-19-808} {\bibfield  {journal}
  {\bibinfo  {journal} {{Quantum}}\ }\textbf {\bibinfo {volume} {6}},\ \bibinfo
  {pages} {808} (\bibinfo {year} {2022})}\BibitemShut {NoStop}%
\bibitem [{\citenamefont {Guan}\ and\ \citenamefont
  {Lewis-Swan}(2021)}]{guan2021}%
  \BibitemOpen
  \bibfield  {author} {\bibinfo {author} {\bibfnamefont {Q.}~\bibnamefont
  {Guan}}\ and\ \bibinfo {author} {\bibfnamefont {R.~J.}\ \bibnamefont
  {Lewis-Swan}},\ }\bibfield  {title} {\bibinfo {title} {Identifying and
  harnessing dynamical phase transitions for quantum-enhanced sensing},\ }\href
  {\doibase 10.1103/PhysRevResearch.3.033199} {\bibfield  {journal} {\bibinfo
  {journal} {Phys. Rev. Res.}\ }\textbf {\bibinfo {volume} {3}},\ \bibinfo
  {pages} {033199} (\bibinfo {year} {2021})}\BibitemShut {NoStop}%
\bibitem [{\citenamefont {Wang}\ and\ \citenamefont
  {Marzolino}(2024)}]{wang2024}%
  \BibitemOpen
  \bibfield  {author} {\bibinfo {author} {\bibfnamefont {Q.}~\bibnamefont
  {Wang}}\ and\ \bibinfo {author} {\bibfnamefont {U.}~\bibnamefont
  {Marzolino}},\ }\bibfield  {title} {\bibinfo {title} {{Precision magnetometry
  exploiting excited state quantum phase transitions}},\ }\href {\doibase
  10.21468/SciPostPhys.17.2.043} {\bibfield  {journal} {\bibinfo  {journal}
  {SciPost Phys.}\ }\textbf {\bibinfo {volume} {17}},\ \bibinfo {pages} {043}
  (\bibinfo {year} {2024})}\BibitemShut {NoStop}%
\bibitem [{\citenamefont {Xue}\ \emph {et~al.}(2026)\citenamefont {Xue},
  \citenamefont {Zhou}, \citenamefont {Huang},\ and\ \citenamefont
  {Li}}]{xue2026}%
  \BibitemOpen
  \bibfield  {author} {\bibinfo {author} {\bibfnamefont {M.}~\bibnamefont
  {Xue}}, \bibinfo {author} {\bibfnamefont {L.}~\bibnamefont {Zhou}}, \bibinfo
  {author} {\bibfnamefont {G.-M.}\ \bibnamefont {Huang}}, \ and\ \bibinfo
  {author} {\bibfnamefont {G.-x.}\ \bibnamefont {Li}},\ }\bibfield  {title}
  {\bibinfo {title} {Quantum fisher information decomposition for optimal
  sensing in the quantum rabi model},\ }\href {\doibase 10.1103/8vkx-b6ds}
  {\bibfield  {journal} {\bibinfo  {journal} {Phys. Rev. A}\ }\textbf {\bibinfo
  {volume} {113}},\ \bibinfo {pages} {013736} (\bibinfo {year}
  {2026})}\BibitemShut {NoStop}%
\bibitem [{\citenamefont {Zhou}\ \emph {et~al.}(2023)\citenamefont {Zhou},
  \citenamefont {Kong}, \citenamefont {Lan},\ and\ \citenamefont
  {Zhang}}]{zhouPRR}%
  \BibitemOpen
  \bibfield  {author} {\bibinfo {author} {\bibfnamefont {L.}~\bibnamefont
  {Zhou}}, \bibinfo {author} {\bibfnamefont {J.}~\bibnamefont {Kong}}, \bibinfo
  {author} {\bibfnamefont {Z.}~\bibnamefont {Lan}}, \ and\ \bibinfo {author}
  {\bibfnamefont {W.}~\bibnamefont {Zhang}},\ }\bibfield  {title} {\bibinfo
  {title} {Dynamical quantum phase transitions in a spinor bose-einstein
  condensate and criticality enhanced quantum sensing},\ }\href {\doibase
  10.1103/PhysRevResearch.5.013087} {\bibfield  {journal} {\bibinfo  {journal}
  {Phys. Rev. Res.}\ }\textbf {\bibinfo {volume} {5}},\ \bibinfo {pages}
  {013087} (\bibinfo {year} {2023})}\BibitemShut {NoStop}%
\bibitem [{\citenamefont {Meyer-Hoppe}\ \emph {et~al.}(2023)\citenamefont
  {Meyer-Hoppe}, \citenamefont {Anders}, \citenamefont {Feldmann},
  \citenamefont {Santos},\ and\ \citenamefont {Klempt}}]{Meyer-Hoppe2023}%
  \BibitemOpen
  \bibfield  {author} {\bibinfo {author} {\bibfnamefont {B.}~\bibnamefont
  {Meyer-Hoppe}}, \bibinfo {author} {\bibfnamefont {F.}~\bibnamefont {Anders}},
  \bibinfo {author} {\bibfnamefont {P.}~\bibnamefont {Feldmann}}, \bibinfo
  {author} {\bibfnamefont {L.}~\bibnamefont {Santos}}, \ and\ \bibinfo {author}
  {\bibfnamefont {C.}~\bibnamefont {Klempt}},\ }\bibfield  {title} {\bibinfo
  {title} {Excited-state phase diagram of a ferromagnetic quantum gas},\ }\href
  {\doibase 10.1103/PhysRevLett.131.243402} {\bibfield  {journal} {\bibinfo
  {journal} {Phys. Rev. Lett.}\ }\textbf {\bibinfo {volume} {131}},\ \bibinfo
  {pages} {243402} (\bibinfo {year} {2023})}\BibitemShut {NoStop}%
\bibitem [{\citenamefont {Austin-Harris}\ \emph {et~al.}()\citenamefont
  {Austin-Harris}, \citenamefont {Sigdel}, \citenamefont {Binegar},
  \citenamefont {Begg}, \citenamefont {Bilitewski},\ and\ \citenamefont
  {Liu}}]{Austin-Harris2025}%
  \BibitemOpen
  \bibfield  {author} {\bibinfo {author} {\bibfnamefont {J.~O.}\ \bibnamefont
  {Austin-Harris}}, \bibinfo {author} {\bibfnamefont {P.}~\bibnamefont
  {Sigdel}}, \bibinfo {author} {\bibfnamefont {C.}~\bibnamefont {Binegar}},
  \bibinfo {author} {\bibfnamefont {S.~E.}\ \bibnamefont {Begg}}, \bibinfo
  {author} {\bibfnamefont {T.}~\bibnamefont {Bilitewski}}, \ and\ \bibinfo
  {author} {\bibfnamefont {Y.}~\bibnamefont {Liu}},\ }\href
  {https://arxiv.org/abs/2511.03720} {\bibinfo {title} {Observation of phase
  memory and dynamical phase transitions in spinor gases},\ }\Eprint
  {http://arxiv.org/abs/2511.03720} {arXiv:2511.03720} \BibitemShut {NoStop}%
\bibitem [{\citenamefont {Lin}\ \emph {et~al.}(2009)\citenamefont {Lin},
  \citenamefont {Compton}, \citenamefont {Jiménez-García}, \citenamefont
  {Porto},\ and\ \citenamefont {Spielman}}]{Lin2009}%
  \BibitemOpen
  \bibfield  {author} {\bibinfo {author} {\bibfnamefont {Y.-J.}\ \bibnamefont
  {Lin}}, \bibinfo {author} {\bibfnamefont {R.~L.}\ \bibnamefont {Compton}},
  \bibinfo {author} {\bibfnamefont {K.}~\bibnamefont {Jiménez-García}},
  \bibinfo {author} {\bibfnamefont {J.~V.}\ \bibnamefont {Porto}}, \ and\
  \bibinfo {author} {\bibfnamefont {I.~B.}\ \bibnamefont {Spielman}},\
  }\bibfield  {title} {\bibinfo {title} {Synthetic magnetic fields for
  ultracold neutral atoms},\ }\href {\doibase 10.1038/nature08609} {\bibfield
  {journal} {\bibinfo  {journal} {Nature}\ }\textbf {\bibinfo {volume} {462}},\
  \bibinfo {pages} {628} (\bibinfo {year} {2009})}\BibitemShut {NoStop}%
\bibitem [{\citenamefont {Lin}\ \emph {et~al.}(2011)\citenamefont {Lin},
  \citenamefont {Jiménez-García},\ and\ \citenamefont {Spielman}}]{Lin2011}%
  \BibitemOpen
  \bibfield  {author} {\bibinfo {author} {\bibfnamefont {Y.-J.}\ \bibnamefont
  {Lin}}, \bibinfo {author} {\bibfnamefont {K.}~\bibnamefont
  {Jiménez-García}}, \ and\ \bibinfo {author} {\bibfnamefont {I.~B.}\
  \bibnamefont {Spielman}},\ }\bibfield  {title} {\bibinfo {title}
  {Spin–orbit-coupled {B}ose–{E}instein condensates},\ }\href {\doibase
  10.1038/nature09887} {\bibfield  {journal} {\bibinfo  {journal} {Nature}\
  }\textbf {\bibinfo {volume} {471}},\ \bibinfo {pages} {83} (\bibinfo {year}
  {2011})}\BibitemShut {NoStop}%
\bibitem [{\citenamefont {Dalibard}\ \emph {et~al.}(2011)\citenamefont
  {Dalibard}, \citenamefont {Gerbier}, \citenamefont
  {Juzeli\ifmmode~\bar{u}\else \={u}\fi{}nas},\ and\ \citenamefont
  {\"Ohberg}}]{Dalibard2011}%
  \BibitemOpen
  \bibfield  {author} {\bibinfo {author} {\bibfnamefont {J.}~\bibnamefont
  {Dalibard}}, \bibinfo {author} {\bibfnamefont {F.}~\bibnamefont {Gerbier}},
  \bibinfo {author} {\bibfnamefont {G.}~\bibnamefont
  {Juzeli\ifmmode~\bar{u}\else \={u}\fi{}nas}}, \ and\ \bibinfo {author}
  {\bibfnamefont {P.}~\bibnamefont {\"Ohberg}},\ }\bibfield  {title} {\bibinfo
  {title} {Colloquium: Artificial gauge potentials for neutral atoms},\ }\href
  {\doibase 10.1103/RevModPhys.83.1523} {\bibfield  {journal} {\bibinfo
  {journal} {Rev. Mod. Phys.}\ }\textbf {\bibinfo {volume} {83}},\ \bibinfo
  {pages} {1523} (\bibinfo {year} {2011})}\BibitemShut {NoStop}%
\bibitem [{\citenamefont {Goldman}\ \emph {et~al.}(2014)\citenamefont
  {Goldman}, \citenamefont {Juzeliūnas}, \citenamefont {Öhberg},\ and\
  \citenamefont {Spielman}}]{Goldman_2014}%
  \BibitemOpen
  \bibfield  {author} {\bibinfo {author} {\bibfnamefont {N.}~\bibnamefont
  {Goldman}}, \bibinfo {author} {\bibfnamefont {G.}~\bibnamefont
  {Juzeliūnas}}, \bibinfo {author} {\bibfnamefont {P.}~\bibnamefont
  {Öhberg}}, \ and\ \bibinfo {author} {\bibfnamefont {I.~B.}\ \bibnamefont
  {Spielman}},\ }\bibfield  {title} {\bibinfo {title} {Light-induced gauge
  fields for ultracold atoms},\ }\href {\doibase
  10.1088/0034-4885/77/12/126401} {\bibfield  {journal} {\bibinfo  {journal}
  {Rep. Prog. Phys.}\ }\textbf {\bibinfo {volume} {77}},\ \bibinfo {pages}
  {126401} (\bibinfo {year} {2014})}\BibitemShut {NoStop}%
\bibitem [{\citenamefont {Zhai}(2015)}]{Zhai_2015}%
  \BibitemOpen
  \bibfield  {author} {\bibinfo {author} {\bibfnamefont {H.}~\bibnamefont
  {Zhai}},\ }\bibfield  {title} {\bibinfo {title} {Degenerate quantum gases
  with spin–orbit coupling: a review},\ }\href {\doibase
  10.1088/0034-4885/78/2/026001} {\bibfield  {journal} {\bibinfo  {journal}
  {Rep. Prog. Phys.}\ }\textbf {\bibinfo {volume} {78}},\ \bibinfo {pages}
  {026001} (\bibinfo {year} {2015})}\BibitemShut {NoStop}%
\bibitem [{\citenamefont {Zhang}\ \emph {et~al.}(2016)\citenamefont {Zhang},
  \citenamefont {Mossman}, \citenamefont {Busch}, \citenamefont {Engels},\ and\
  \citenamefont {Zhang}}]{Zhang2016}%
  \BibitemOpen
  \bibfield  {author} {\bibinfo {author} {\bibfnamefont {Y.}~\bibnamefont
  {Zhang}}, \bibinfo {author} {\bibfnamefont {M.~E.}\ \bibnamefont {Mossman}},
  \bibinfo {author} {\bibfnamefont {T.}~\bibnamefont {Busch}}, \bibinfo
  {author} {\bibfnamefont {P.}~\bibnamefont {Engels}}, \ and\ \bibinfo {author}
  {\bibfnamefont {C.}~\bibnamefont {Zhang}},\ }\bibfield  {title} {\bibinfo
  {title} {Properties of spin--orbit-coupled bose--einstein condensates},\
  }\href {\doibase 10.1007/s11467-016-0560-y} {\bibfield  {journal} {\bibinfo
  {journal} {Front. Phys.}\ }\textbf {\bibinfo {volume} {11}},\ \bibinfo
  {pages} {118103} (\bibinfo {year} {2016})}\BibitemShut {NoStop}%
\bibitem [{\citenamefont {Anderson}\ \emph {et~al.}(2011)\citenamefont
  {Anderson}, \citenamefont {Taylor},\ and\ \citenamefont
  {Galitski}}]{Anderson2011}%
  \BibitemOpen
  \bibfield  {author} {\bibinfo {author} {\bibfnamefont {B.~M.}\ \bibnamefont
  {Anderson}}, \bibinfo {author} {\bibfnamefont {J.~M.}\ \bibnamefont
  {Taylor}}, \ and\ \bibinfo {author} {\bibfnamefont {V.~M.}\ \bibnamefont
  {Galitski}},\ }\bibfield  {title} {\bibinfo {title} {Interferometry with
  synthetic gauge fields},\ }\href {\doibase 10.1103/PhysRevA.83.031602}
  {\bibfield  {journal} {\bibinfo  {journal} {Phys. Rev. A}\ }\textbf {\bibinfo
  {volume} {83}},\ \bibinfo {pages} {031602} (\bibinfo {year}
  {2011})}\BibitemShut {NoStop}%
\bibitem [{\citenamefont {Jacob}\ \emph {et~al.}(2007)\citenamefont {Jacob},
  \citenamefont {{\"O}hberg}, \citenamefont {Juzeli{\={u}}nas},\ and\
  \citenamefont {Santos}}]{Jacob2007}%
  \BibitemOpen
  \bibfield  {author} {\bibinfo {author} {\bibfnamefont {A.}~\bibnamefont
  {Jacob}}, \bibinfo {author} {\bibfnamefont {P.}~\bibnamefont {{\"O}hberg}},
  \bibinfo {author} {\bibfnamefont {G.}~\bibnamefont {Juzeli{\={u}}nas}}, \
  and\ \bibinfo {author} {\bibfnamefont {L.}~\bibnamefont {Santos}},\
  }\bibfield  {title} {\bibinfo {title} {Cold atom dynamics in non-abelian
  gauge fields},\ }\href {\doibase 10.1007/s00340-007-2865-6} {\bibfield
  {journal} {\bibinfo  {journal} {Appl. Phys. B}\ }\textbf {\bibinfo {volume}
  {89}},\ \bibinfo {pages} {439} (\bibinfo {year} {2007})}\BibitemShut
  {NoStop}%
\bibitem [{\citenamefont {Osterloh}\ \emph {et~al.}(2005)\citenamefont
  {Osterloh}, \citenamefont {Baig}, \citenamefont {Santos}, \citenamefont
  {Zoller},\ and\ \citenamefont {Lewenstein}}]{Osterloh2005}%
  \BibitemOpen
  \bibfield  {author} {\bibinfo {author} {\bibfnamefont {K.}~\bibnamefont
  {Osterloh}}, \bibinfo {author} {\bibfnamefont {M.}~\bibnamefont {Baig}},
  \bibinfo {author} {\bibfnamefont {L.}~\bibnamefont {Santos}}, \bibinfo
  {author} {\bibfnamefont {P.}~\bibnamefont {Zoller}}, \ and\ \bibinfo {author}
  {\bibfnamefont {M.}~\bibnamefont {Lewenstein}},\ }\bibfield  {title}
  {\bibinfo {title} {Cold atoms in non-abelian gauge potentials: From the
  hofstadter "moth" to lattice gauge theory},\ }\href {\doibase
  10.1103/PhysRevLett.95.010403} {\bibfield  {journal} {\bibinfo  {journal}
  {Phys. Rev. Lett.}\ }\textbf {\bibinfo {volume} {95}},\ \bibinfo {pages}
  {010403} (\bibinfo {year} {2005})}\BibitemShut {NoStop}%
\bibitem [{\citenamefont {Olson}\ \emph {et~al.}(2017)\citenamefont {Olson},
  \citenamefont {Blasing}, \citenamefont {Qu}, \citenamefont {Li},
  \citenamefont {Niffenegger}, \citenamefont {Zhang},\ and\ \citenamefont
  {Chen}}]{Olson2017}%
  \BibitemOpen
  \bibfield  {author} {\bibinfo {author} {\bibfnamefont {A.~J.}\ \bibnamefont
  {Olson}}, \bibinfo {author} {\bibfnamefont {D.~B.}\ \bibnamefont {Blasing}},
  \bibinfo {author} {\bibfnamefont {C.}~\bibnamefont {Qu}}, \bibinfo {author}
  {\bibfnamefont {C.-H.}\ \bibnamefont {Li}}, \bibinfo {author} {\bibfnamefont
  {R.~J.}\ \bibnamefont {Niffenegger}}, \bibinfo {author} {\bibfnamefont
  {C.}~\bibnamefont {Zhang}}, \ and\ \bibinfo {author} {\bibfnamefont {Y.~P.}\
  \bibnamefont {Chen}},\ }\bibfield  {title} {\bibinfo {title} {Stueckelberg
  interferometry using periodically driven spin-orbit-coupled bose-einstein
  condensates},\ }\href {\doibase 10.1103/PhysRevA.95.043623} {\bibfield
  {journal} {\bibinfo  {journal} {Phys. Rev. A}\ }\textbf {\bibinfo {volume}
  {95}},\ \bibinfo {pages} {043623} (\bibinfo {year} {2017})}\BibitemShut
  {NoStop}%
\bibitem [{\citenamefont {Liang}\ \emph {et~al.}(2020)\citenamefont {Liang},
  \citenamefont {Li}, \citenamefont {Zhang}, \citenamefont {Zhou},\ and\
  \citenamefont {Lan}}]{Liang2020}%
  \BibitemOpen
  \bibfield  {author} {\bibinfo {author} {\bibfnamefont {S.}~\bibnamefont
  {Liang}}, \bibinfo {author} {\bibfnamefont {Z.-C.}\ \bibnamefont {Li}},
  \bibinfo {author} {\bibfnamefont {W.}~\bibnamefont {Zhang}}, \bibinfo
  {author} {\bibfnamefont {L.}~\bibnamefont {Zhou}}, \ and\ \bibinfo {author}
  {\bibfnamefont {Z.}~\bibnamefont {Lan}},\ }\bibfield  {title} {\bibinfo
  {title} {St\"uckelberg interferometry using spin-orbit-coupled cold atoms in
  an optical lattice},\ }\href {\doibase 10.1103/PhysRevA.102.033332}
  {\bibfield  {journal} {\bibinfo  {journal} {Phys. Rev. A}\ }\textbf {\bibinfo
  {volume} {102}},\ \bibinfo {pages} {033332} (\bibinfo {year}
  {2020})}\BibitemShut {NoStop}%
\bibitem [{\citenamefont {Gietka}\ and\ \citenamefont
  {Ritsch}(2023)}]{Gietka2023}%
  \BibitemOpen
  \bibfield  {author} {\bibinfo {author} {\bibfnamefont {K.}~\bibnamefont
  {Gietka}}\ and\ \bibinfo {author} {\bibfnamefont {H.}~\bibnamefont
  {Ritsch}},\ }\bibfield  {title} {\bibinfo {title} {Squeezing and overcoming
  the heisenberg scaling with spin-orbit coupled quantum gases},\ }\href
  {\doibase 10.1103/PhysRevLett.130.090802} {\bibfield  {journal} {\bibinfo
  {journal} {Phys. Rev. Lett.}\ }\textbf {\bibinfo {volume} {130}},\ \bibinfo
  {pages} {090802} (\bibinfo {year} {2023})}\BibitemShut {NoStop}%
\bibitem [{\citenamefont {Cabedo}\ \emph {et~al.}(2021)\citenamefont {Cabedo},
  \citenamefont {Claramunt},\ and\ \citenamefont {Celi}}]{CabedoPRA2021}%
  \BibitemOpen
  \bibfield  {author} {\bibinfo {author} {\bibfnamefont {J.}~\bibnamefont
  {Cabedo}}, \bibinfo {author} {\bibfnamefont {J.}~\bibnamefont {Claramunt}}, \
  and\ \bibinfo {author} {\bibfnamefont {A.}~\bibnamefont {Celi}},\ }\bibfield
  {title} {\bibinfo {title} {Dynamical preparation of stripe states in
  spin-orbit-coupled gases},\ }\href {\doibase 10.1103/PhysRevA.104.L031305}
  {\bibfield  {journal} {\bibinfo  {journal} {Phys. Rev. A}\ }\textbf {\bibinfo
  {volume} {104}},\ \bibinfo {pages} {L031305} (\bibinfo {year}
  {2021})}\BibitemShut {NoStop}%
\bibitem [{\citenamefont {Cabedo}\ and\ \citenamefont
  {Celi}(2021)}]{Cabedo2021}%
  \BibitemOpen
  \bibfield  {author} {\bibinfo {author} {\bibfnamefont {J.}~\bibnamefont
  {Cabedo}}\ and\ \bibinfo {author} {\bibfnamefont {A.}~\bibnamefont {Celi}},\
  }\bibfield  {title} {\bibinfo {title} {Excited-state quantum phase
  transitions in spin-orbit-coupled bose gases},\ }\href {\doibase
  10.1103/PhysRevResearch.3.043215} {\bibfield  {journal} {\bibinfo  {journal}
  {Phys. Rev. Res.}\ }\textbf {\bibinfo {volume} {3}},\ \bibinfo {pages}
  {043215} (\bibinfo {year} {2021})}\BibitemShut {NoStop}%
\bibitem [{\citenamefont {Lan}\ and\ \citenamefont {\"Ohberg}(2014)}]{lan2014}%
  \BibitemOpen
  \bibfield  {author} {\bibinfo {author} {\bibfnamefont {Z.}~\bibnamefont
  {Lan}}\ and\ \bibinfo {author} {\bibfnamefont {P.}~\bibnamefont {\"Ohberg}},\
  }\bibfield  {title} {\bibinfo {title} {Raman-dressed spin-1
  spin-orbit-coupled quantum gas},\ }\href {\doibase
  10.1103/PhysRevA.89.023630} {\bibfield  {journal} {\bibinfo  {journal} {Phys.
  Rev. A}\ }\textbf {\bibinfo {volume} {89}},\ \bibinfo {pages} {023630}
  (\bibinfo {year} {2014})}\BibitemShut {NoStop}%
\bibitem [{\citenamefont {Campbell}\ \emph {et~al.}(2016)\citenamefont
  {Campbell}, \citenamefont {Price}, \citenamefont {Putra}, \citenamefont
  {Vald{\'e}s-Curiel}, \citenamefont {Trypogeorgos},\ and\ \citenamefont
  {Spielman}}]{Campbell2016}%
  \BibitemOpen
  \bibfield  {author} {\bibinfo {author} {\bibfnamefont {D.~L.}\ \bibnamefont
  {Campbell}}, \bibinfo {author} {\bibfnamefont {R.~M.}\ \bibnamefont {Price}},
  \bibinfo {author} {\bibfnamefont {A.}~\bibnamefont {Putra}}, \bibinfo
  {author} {\bibfnamefont {A.}~\bibnamefont {Vald{\'e}s-Curiel}}, \bibinfo
  {author} {\bibfnamefont {D.}~\bibnamefont {Trypogeorgos}}, \ and\ \bibinfo
  {author} {\bibfnamefont {I.~B.}\ \bibnamefont {Spielman}},\ }\bibfield
  {title} {\bibinfo {title} {Magnetic phases of spin-1 spin--orbit-coupled bose
  gases},\ }\href {\doibase 10.1038/ncomms10897} {\bibfield  {journal}
  {\bibinfo  {journal} {Nat. Commun.}\ }\textbf {\bibinfo {volume} {7}},\
  \bibinfo {pages} {10897} (\bibinfo {year} {2016})}\BibitemShut {NoStop}%
\bibitem [{\citenamefont {Gerbier}\ \emph {et~al.}(2006)\citenamefont
  {Gerbier}, \citenamefont {Widera}, \citenamefont {F\"olling}, \citenamefont
  {Mandel},\ and\ \citenamefont {Bloch}}]{Gerbier2006}%
  \BibitemOpen
  \bibfield  {author} {\bibinfo {author} {\bibfnamefont {F.}~\bibnamefont
  {Gerbier}}, \bibinfo {author} {\bibfnamefont {A.}~\bibnamefont {Widera}},
  \bibinfo {author} {\bibfnamefont {S.}~\bibnamefont {F\"olling}}, \bibinfo
  {author} {\bibfnamefont {O.}~\bibnamefont {Mandel}}, \ and\ \bibinfo {author}
  {\bibfnamefont {I.}~\bibnamefont {Bloch}},\ }\bibfield  {title} {\bibinfo
  {title} {Resonant control of spin dynamics in ultracold quantum gases by
  microwave dressing},\ }\href {\doibase 10.1103/PhysRevA.73.041602} {\bibfield
   {journal} {\bibinfo  {journal} {Phys. Rev. A}\ }\textbf {\bibinfo {volume}
  {73}},\ \bibinfo {pages} {041602} (\bibinfo {year} {2006})}\BibitemShut
  {NoStop}%
\bibitem [{\citenamefont {Leslie}\ \emph {et~al.}(2009)\citenamefont {Leslie},
  \citenamefont {Guzman}, \citenamefont {Vengalattore}, \citenamefont {Sau},
  \citenamefont {Cohen},\ and\ \citenamefont {Stamper-Kurn}}]{Leslie2009}%
  \BibitemOpen
  \bibfield  {author} {\bibinfo {author} {\bibfnamefont {S.~R.}\ \bibnamefont
  {Leslie}}, \bibinfo {author} {\bibfnamefont {J.}~\bibnamefont {Guzman}},
  \bibinfo {author} {\bibfnamefont {M.}~\bibnamefont {Vengalattore}}, \bibinfo
  {author} {\bibfnamefont {J.~D.}\ \bibnamefont {Sau}}, \bibinfo {author}
  {\bibfnamefont {M.~L.}\ \bibnamefont {Cohen}}, \ and\ \bibinfo {author}
  {\bibfnamefont {D.~M.}\ \bibnamefont {Stamper-Kurn}},\ }\bibfield  {title}
  {\bibinfo {title} {Amplification of fluctuations in a spinor bose-einstein
  condensate},\ }\href {\doibase 10.1103/PhysRevA.79.043631} {\bibfield
  {journal} {\bibinfo  {journal} {Phys. Rev. A}\ }\textbf {\bibinfo {volume}
  {79}},\ \bibinfo {pages} {043631} (\bibinfo {year} {2009})}\BibitemShut
  {NoStop}%
\bibitem [{\citenamefont {Sadler}\ \emph {et~al.}(2006)\citenamefont {Sadler},
  \citenamefont {Higbie}, \citenamefont {Leslie}, \citenamefont
  {Vengalattore},\ and\ \citenamefont {Stamper-Kurn}}]{Sadler2006}%
  \BibitemOpen
  \bibfield  {author} {\bibinfo {author} {\bibfnamefont {L.~E.}\ \bibnamefont
  {Sadler}}, \bibinfo {author} {\bibfnamefont {J.~M.}\ \bibnamefont {Higbie}},
  \bibinfo {author} {\bibfnamefont {S.~R.}\ \bibnamefont {Leslie}}, \bibinfo
  {author} {\bibfnamefont {M.}~\bibnamefont {Vengalattore}}, \ and\ \bibinfo
  {author} {\bibfnamefont {D.~M.}\ \bibnamefont {Stamper-Kurn}},\ }\bibfield
  {title} {\bibinfo {title} {Spontaneous symmetry breaking in a quenched
  ferromagnetic spinor bose--einstein condensate},\ }\href {\doibase
  10.1038/nature05094} {\bibfield  {journal} {\bibinfo  {journal} {Nature}\
  }\textbf {\bibinfo {volume} {443}},\ \bibinfo {pages} {312} (\bibinfo {year}
  {2006})}\BibitemShut {NoStop}%
\bibitem [{\citenamefont {Murata}\ \emph {et~al.}(2007)\citenamefont {Murata},
  \citenamefont {Saito},\ and\ \citenamefont {Ueda}}]{Murata2007}%
  \BibitemOpen
  \bibfield  {author} {\bibinfo {author} {\bibfnamefont {K.}~\bibnamefont
  {Murata}}, \bibinfo {author} {\bibfnamefont {H.}~\bibnamefont {Saito}}, \
  and\ \bibinfo {author} {\bibfnamefont {M.}~\bibnamefont {Ueda}},\ }\bibfield
  {title} {\bibinfo {title} {Broken-axisymmetry phase of a spin-1 ferromagnetic
  bose-einstein condensate},\ }\href {\doibase 10.1103/PhysRevA.75.013607}
  {\bibfield  {journal} {\bibinfo  {journal} {Phys. Rev. A}\ }\textbf {\bibinfo
  {volume} {75}},\ \bibinfo {pages} {013607} (\bibinfo {year}
  {2007})}\BibitemShut {NoStop}%
\bibitem [{\citenamefont {Davis}\ \emph {et~al.}(2016)\citenamefont {Davis},
  \citenamefont {Bentsen},\ and\ \citenamefont {Schleier-Smith}}]{Davis2016}%
  \BibitemOpen
  \bibfield  {author} {\bibinfo {author} {\bibfnamefont {E.}~\bibnamefont
  {Davis}}, \bibinfo {author} {\bibfnamefont {G.}~\bibnamefont {Bentsen}}, \
  and\ \bibinfo {author} {\bibfnamefont {M.}~\bibnamefont {Schleier-Smith}},\
  }\bibfield  {title} {\bibinfo {title} {Approaching the heisenberg limit
  without single-particle detection},\ }\href {\doibase
  10.1103/PhysRevLett.116.053601} {\bibfield  {journal} {\bibinfo  {journal}
  {Phys. Rev. Lett.}\ }\textbf {\bibinfo {volume} {116}},\ \bibinfo {pages}
  {053601} (\bibinfo {year} {2016})}\BibitemShut {NoStop}%
\bibitem [{\citenamefont {Fr\"owis}\ \emph {et~al.}(2016)\citenamefont
  {Fr\"owis}, \citenamefont {Sekatski},\ and\ \citenamefont
  {D\"ur}}]{Florian2016}%
  \BibitemOpen
  \bibfield  {author} {\bibinfo {author} {\bibfnamefont {F.}~\bibnamefont
  {Fr\"owis}}, \bibinfo {author} {\bibfnamefont {P.}~\bibnamefont {Sekatski}},
  \ and\ \bibinfo {author} {\bibfnamefont {W.}~\bibnamefont {D\"ur}},\
  }\bibfield  {title} {\bibinfo {title} {Detecting large quantum fisher
  information with finite measurement precision},\ }\href {\doibase
  10.1103/PhysRevLett.116.090801} {\bibfield  {journal} {\bibinfo  {journal}
  {Phys. Rev. Lett.}\ }\textbf {\bibinfo {volume} {116}},\ \bibinfo {pages}
  {090801} (\bibinfo {year} {2016})}\BibitemShut {NoStop}%
\bibitem [{\citenamefont {Macr\`{\i}}\ \emph {et~al.}(2016)\citenamefont
  {Macr\`{\i}}, \citenamefont {Smerzi},\ and\ \citenamefont
  {Pezz\`e}}]{Tommaso2016}%
  \BibitemOpen
  \bibfield  {author} {\bibinfo {author} {\bibfnamefont {T.}~\bibnamefont
  {Macr\`{\i}}}, \bibinfo {author} {\bibfnamefont {A.}~\bibnamefont {Smerzi}},
  \ and\ \bibinfo {author} {\bibfnamefont {L.}~\bibnamefont {Pezz\`e}},\
  }\bibfield  {title} {\bibinfo {title} {Loschmidt echo for quantum
  metrology},\ }\href {\doibase 10.1103/PhysRevA.94.010102} {\bibfield
  {journal} {\bibinfo  {journal} {Phys. Rev. A}\ }\textbf {\bibinfo {volume}
  {94}},\ \bibinfo {pages} {010102(R)} (\bibinfo {year} {2016})}\BibitemShut
  {NoStop}%
\bibitem [{\citenamefont {Niezgoda}\ \emph {et~al.}(2019)\citenamefont
  {Niezgoda}, \citenamefont {Kajtoch}, \citenamefont {Dziekańska},\ and\
  \citenamefont {Witkowska}}]{Niezgoda_2019}%
  \BibitemOpen
  \bibfield  {author} {\bibinfo {author} {\bibfnamefont {A.}~\bibnamefont
  {Niezgoda}}, \bibinfo {author} {\bibfnamefont {D.}~\bibnamefont {Kajtoch}},
  \bibinfo {author} {\bibfnamefont {J.}~\bibnamefont {Dziekańska}}, \ and\
  \bibinfo {author} {\bibfnamefont {E.}~\bibnamefont {Witkowska}},\ }\bibfield
  {title} {\bibinfo {title} {Optimal quantum interferometry robust to detection
  noise using spin-1 atomic condensates},\ }\href {\doibase
  10.1088/1367-2630/ab4099} {\bibfield  {journal} {\bibinfo  {journal} {New J.
  Phys.}\ }\textbf {\bibinfo {volume} {21}},\ \bibinfo {pages} {093037}
  (\bibinfo {year} {2019})}\BibitemShut {NoStop}%
\bibitem [{\citenamefont {Chelpanova}\ \emph {et~al.}(2024)\citenamefont
  {Chelpanova}, \citenamefont {Seetharam}, \citenamefont {Rosa-Medina},
  \citenamefont {Reiter}, \citenamefont {Finger}, \citenamefont {Donner},\ and\
  \citenamefont {Marino}}]{Oksana2024}%
  \BibitemOpen
  \bibfield  {author} {\bibinfo {author} {\bibfnamefont {O.}~\bibnamefont
  {Chelpanova}}, \bibinfo {author} {\bibfnamefont {K.}~\bibnamefont
  {Seetharam}}, \bibinfo {author} {\bibfnamefont {R.}~\bibnamefont
  {Rosa-Medina}}, \bibinfo {author} {\bibfnamefont {N.}~\bibnamefont {Reiter}},
  \bibinfo {author} {\bibfnamefont {F.}~\bibnamefont {Finger}}, \bibinfo
  {author} {\bibfnamefont {T.}~\bibnamefont {Donner}}, \ and\ \bibinfo {author}
  {\bibfnamefont {J.}~\bibnamefont {Marino}},\ }\bibfield  {title} {\bibinfo
  {title} {Dynamics of spin-momentum entanglement from superradiant phase
  transitions},\ }\href {\doibase 10.1103/PhysRevResearch.6.033193} {\bibfield
  {journal} {\bibinfo  {journal} {Phys. Rev. Res.}\ }\textbf {\bibinfo {volume}
  {6}},\ \bibinfo {pages} {033193} (\bibinfo {year} {2024})}\BibitemShut
  {NoStop}%
\bibitem [{\citenamefont {Finger}\ \emph {et~al.}(2024)\citenamefont {Finger},
  \citenamefont {Rosa-Medina}, \citenamefont {Reiter}, \citenamefont
  {Christodoulou}, \citenamefont {Donner},\ and\ \citenamefont
  {Esslinger}}]{Finger2024}%
  \BibitemOpen
  \bibfield  {author} {\bibinfo {author} {\bibfnamefont {F.}~\bibnamefont
  {Finger}}, \bibinfo {author} {\bibfnamefont {R.}~\bibnamefont {Rosa-Medina}},
  \bibinfo {author} {\bibfnamefont {N.}~\bibnamefont {Reiter}}, \bibinfo
  {author} {\bibfnamefont {P.}~\bibnamefont {Christodoulou}}, \bibinfo {author}
  {\bibfnamefont {T.}~\bibnamefont {Donner}}, \ and\ \bibinfo {author}
  {\bibfnamefont {T.}~\bibnamefont {Esslinger}},\ }\bibfield  {title} {\bibinfo
  {title} {Spin- and momentum-correlated atom pairs mediated by photon exchange
  and seeded by vacuum fluctuations},\ }\href {\doibase
  10.1103/PhysRevLett.132.093402} {\bibfield  {journal} {\bibinfo  {journal}
  {Phys. Rev. Lett.}\ }\textbf {\bibinfo {volume} {132}},\ \bibinfo {pages}
  {093402} (\bibinfo {year} {2024})}\BibitemShut {NoStop}%
\bibitem [{\citenamefont {Ravisankar}\ \emph {et~al.}(2021)\citenamefont
  {Ravisankar}, \citenamefont {Vudragović}, \citenamefont {Muruganandam},
  \citenamefont {Balaž},\ and\ \citenamefont
  {Adhikari}}]{RAVISANKAR2021107657}%
  \BibitemOpen
  \bibfield  {author} {\bibinfo {author} {\bibfnamefont {R.}~\bibnamefont
  {Ravisankar}}, \bibinfo {author} {\bibfnamefont {D.}~\bibnamefont
  {Vudragović}}, \bibinfo {author} {\bibfnamefont {P.}~\bibnamefont
  {Muruganandam}}, \bibinfo {author} {\bibfnamefont {A.}~\bibnamefont
  {Balaž}}, \ and\ \bibinfo {author} {\bibfnamefont {S.~K.}\ \bibnamefont
  {Adhikari}},\ }\bibfield  {title} {\bibinfo {title} {Spin-1 spin–orbit- and
  rabi-coupled bose–einstein condensate solver},\ }\href {\doibase
  https://doi.org/10.1016/j.cpc.2020.107657} {\bibfield  {journal} {\bibinfo
  {journal} {Comput. Phys. Commun.}\ }\textbf {\bibinfo {volume} {259}},\
  \bibinfo {pages} {107657} (\bibinfo {year} {2021})}\BibitemShut {NoStop}%
\end{thebibliography}
%merlin.mbs apsrev4-1.bst 2010-07-25 4.21a (PWD, AO, DPC) hacked
%Control: key (0)
%Control: author (8) initials jnrlst
%Control: editor formatted (1) identically to author
%Control: production of article title (-1) disabled
%Control: page (0) single
%Control: year (1) truncated
%Control: production of eprint (0) enabled
%

\end{document}